\RequirePackage{fix-cm}
\documentclass[smallextended]{svjour3}       
\usepackage{amsmath}
\usepackage{enumitem}
\usepackage{hyperref}
\usepackage{indentfirst}

\smartqed  

\usepackage{graphicx}
\usepackage{csquotes}
\usepackage{threeparttable}
\usepackage{booktabs} 
\usepackage{amsmath}
\usepackage{longtable} 

\usepackage{color, colortbl}
\usepackage[many]{tcolorbox}    	
\usepackage{hyperref}
\usepackage{multirow}

\usepackage{xcolor}
\normalmarginpar

\begin{document}

\title{Reducing Labeling Effort in Architecture Technical Debt Detection through Active Learning and Explainable AI}

\author{Edi Sutoyo \and
        Paris Avgeriou \and 
        Andrea Capiluppi
} 
\authorrunning{E. Sutoyo et al.}
%
\institute{E. Sutoyo \at 
            1. Bernoulli Institute, University of Groningen, Groningen, The Netherlands\\
            2. Department of Information Systems, Telkom University, Bandung, Indonesia\\
            \email{e.sutoyo@rug.nl}
            \and
            P. Avgeriou \at
            Bernoulli Institute, University of Groningen, Groningen, The Netherlands\\
            \email{p.avgeriou@rug.nl}
            \and
            A. Capiluppi \at
            Bernoulli Institute, University of Groningen, Groningen, The Netherlands\\
            \email{a.capiluppi@rug.nl}
            }


%

\date{Received: date / Accepted: date}

\maketitle              
%
\begin{abstract}
Self{-}Admitted Technical Debt (SATD) refers to technical compromises explicitly admitted by developers in natural language artifacts, such as code comments, commit messages, and issue trackers. Among its types, Architecture Technical Debt (ATD) is particularly difficult to detect due to its abstract and context-dependent nature. Manual annotation of ATD is costly, time-consuming, and challenging to scale. To reduce labeling effort, this study combines keyword-based filtering, active learning, and explainable AI for ATD detection. We refined an existing dataset of ATD{-}related Jira issues to obtain an expert{-}validated seed set used to extract representative keywords. These keywords were then applied to identify more than 103k candidate issues across 10 open-source projects. To assess the reliability of keyword-based filtering, we qualitatively evaluated a statistically representative sample of labeled issues. Building on the resulting dataset, we applied active learning with multiple query strategies to prioritize informative samples for annotation. The results show that Breaking Ties achieved the best performance, with an F1-score of 0.72 and a 49\% reduction in annotation effort. To improve transparency, we used SHAP and LIME to explain ATD classification results. Expert evaluation showed that both methods provided useful explanations, with LIME generally preferred for its clarity and ease of use.

\end{abstract}



\section{Introduction}
\label{intro}
Technical debt (TD) represents the accumulated compromises, shortcuts, and suboptimal decisions made during development that can impede system evolution and quality over time \cite{cunningham1992wycash}. Among its various forms, \textit{architecture technical debt} (ATD) is particularly critical because it affects the system at the architectural level and can have broad, long-lasting consequences \cite{martini2015danger}. ATD emerges when architectural decisions prioritize short-term development needs over long-term quality concerns. Typical examples include introducing cyclic dependencies among modules or bypassing established architectural layers to accelerate development \cite{nayebi2019longitudinal}. Unlike code-level debt, ATD often spans multiple components, making it more difficult and costly to identify and repay \cite{carrillo2018ripple,xiao2016identifying}.

Despite its significance, detecting ATD remains challenging because architectural decisions are often complex and abstract. As a result, conventional automated approaches, such as static analysis or generic repository mining, often fail to capture the subtle linguistic and conceptual signals associated with ATD. A complementary direction is the detection of \textit{Self{-}Admitted Technical Debt} (SATD), where developers explicitly acknowledge technical limitations, trade-offs, or compromises in natural language artifacts such as code comments, commit messages, pull requests, and issue reports~\cite{dai2017detecting,codabux2021technical,li2022identifying,sheikhaei2024empirical}. SATD is particularly relevant because, compared with TD identified using code metrics or code smells, it is generally considered more reliable, as developers explicitly admit it~\cite {da2017using}. In contrast, TD identified through code metrics or code smells may suffer from high false-positive rates~\cite{fontana2016antipattern}. SATD identification is also more lightweight than source code analysis-based TD detection because it does not require the construction of abstract syntax trees or other computationally expensive program representations~\cite{da2017using}. This is particularly relevant for ATD, as developers often discuss design limitations, dependency issues, migration needs, and obsolete technologies in issue tracking systems \cite{li2022identifying}.

Prior research has shown that ATD is an important and underexplored type of debt, and several studies have examined its nature, consequences, and evolution~\cite{lenarduzzi2019technical,martini2014architecture}. In this study, we treat ATD expressed in issue reports as a specific type of SATD. While SATD concerns how technical debt is explicitly expressed in textual artifacts, ATD refers to the architectural nature of the underlying problem, namely, debt that affects system-level structure, such as problematic dependencies, modularity violations, architectural erosion, or the continued use of architecturally significant obsolete technologies~\cite{li2023automatic}. Accordingly, this study focuses on issue reports in which developers explicitly describe architecture-related debt.

This distinction is important because ATD has consistently been identified as the most harmful and costly type of TD~\cite{lenarduzzi2019technical}, with architectural issues often causing widespread and persistent impacts on system maintainability, evolution, and cost~\cite{martini2014architecture}. Moreover, both researchers and practitioners frequently identify architectural debt as the type of technical debt that requires the most management attention~\cite{lenarduzzi2019technical}. Dedicated ATD detection is therefore practically important, as identifying debt-related text alone is often insufficient for practitioners who must distinguish architectural concerns from other forms of debt to prioritize issues and assign appropriate remediation actions. Despite this need, most SATD studies have focused on generic forms of TD (e.g., code/design debt, requirement debt, documentation debt, and test debt)~\cite{li2023automatic,sutoyo2024deep}, while explicit SATD related to ATD remains underexplored~\cite{sutoyo2026self}.

Existing ATD studies also generally assume that ATD items have already been identified. In contrast, the earlier problem of how to efficiently identify and label ATD in issue reports remains underexplored. Addressing this gap is important because both empirical ATD studies and machine learning models depend on reliable annotated datasets, yet constructing such datasets requires costly expert effort.


Active learning offers a promising way to address this bottleneck. Although active learning has been applied in several software engineering (SE) tasks \cite{bowring2004active}, its application to ATD detection poses a distinct challenge. ATD labels are scarce, require architectural expertise, and are often expressed only indirectly in issue reports. This makes ATD detection a particularly suitable, but also demanding, setting for active learning, where the goal is not only to improve predictive performance but also to reduce the cost of constructing reliable labeled data for an underrepresented architectural debt category.

In addition, accurate prediction alone is not sufficient for practical use. For ATD detection to support SE practice, users need to understand why an issue is classified as ATD. This makes explainable AI (XAI) particularly relevant, since interpretable explanations can help experts assess whether model decisions are meaningful and trustworthy in an architectural context~\cite{tantithamthavorn2021explainable}. The importance of such transparency is also highlighted in the recent Dagstuhl Perspectives Workshop Manifesto~\cite{avgeriou2025manifesto}, which advocates TD identification approaches that are directly traceable to specific data sources and explainable to all stakeholders.


To address these challenges, we propose a hybrid pipeline to reduce labeling effort for ATD detection in Jira issue reports. Starting from 57 expert{-}validated seed issues refined from 116 ATD{-}related issues in prior work, we extract representative ATD keywords using multiple techniques. We then use these keywords to retrieve issues from ten large-scale open-source projects and manually validate statistically representative samples to construct a labeled dataset. Building on this dataset, we apply active learning with multiple query strategies and transformer-based models to improve classification while reducing annotation effort. Finally, we use LIME and SHAP to explain model predictions and assess their usefulness through expert evaluation.


The key contributions of this study are as follows.

\begin{itemize}
    \item \textbf{A hybrid pipeline and novel dataset for ATD identification from Jira issue reports.} We combine dataset refinement, keyword extraction, issue filtering, manual validation, and supervised classification to support scalable ATD dataset construction. In doing so, we contribute a novel dataset of ATD in issue tracking systems.

    \item \textbf{An active learning strategy for reducing ATD labeling effort.} We investigate active learning in the specific context of ATD, where labels are scarce, expert-dependent, and costly to obtain, and show that uncertainty-based sampling can improve detection performance.

    \item \textbf{An explainable and expert-evaluated ATD detection framework.} We integrate LIME and SHAP into the ATD classification pipeline and assess their usefulness through expert evaluation. To the best of our knowledge, this is the first work to systematically involve domain experts in evaluating XAI outputs for text-based ATD detection.
\end{itemize}



The remainder of this paper is structured as follows. Section~\ref{sectionBG} reviews the background and related work. Section~\ref{sectionRM} presents the study design and methodology. Section~\ref{sectionResults} reports the results. Section~\ref{sectionDiscussion} discusses the findings, implications, and threats to validity. Section~\ref{sectionConclusion} concludes the study.




\section{Background and Related Work}
\label{sectionBG}

In this section, we review the literature relevant to our study in three areas: (i) research on Architecture Technical Debt, including self{-}admitted ATD; (ii) applications of active learning in SE; and (iii) the use of explainable AI in SE.

\subsection{Architecture Technical Debt}
ATD is commonly studied through structural and architectural analysis techniques. Unlike code-level debt, ATD often propagates across multiple components and design decisions, making it more difficult and costly to detect and repay~\cite{carrillo2018ripple,xiao2016identifying}. Prior work has used architectural smells, dependency analysis, coupling metrics, modularity violations, and other architecture recovery or conformance checking techniques to identify ATD symptoms in software systems~\cite{martini2015investigating,xiao2016identifying,nayebi2019longitudinal,verdecchia2020architectural}. However, these techniques also have limitations. Architectural smells or structural anomalies do not always correspond to actual debt, as some may reflect intentional design trade-offs, temporary adaptations, or context-dependent decisions. As a result, structural detection approaches may suffer from false positives and may not fully reveal whether developers themselves perceive a given architectural problem as debt~\cite{fontana2016antipattern}.

To complement these structural approaches, researchers have also studied SATD, which refers to cases where developers explicitly acknowledge technical shortcomings, workarounds, or design compromises in natural language artifacts such as code comments, commit messages, and issue reports~\cite{sutoyo2026self}. Because SATD is explicitly documented by practitioners, it can provide direct evidence of perceived debt and its rationale. Early SATD research focused on source code comments, with Potdar and Shihab~\cite{potdar2014exploratory} identifying 62 SATD patterns and Maldonado et al.~\cite{da2017using} later showing that supervised NLP methods can detect SATD more effectively than keyword-based approaches.

In this study, we treat ATD as a specific debt type of SATD. SATD broadly captures developer-admitted technical compromises in textual artifacts, whereas ATD refers specifically to those with system-level structural implications, such as modularity violations, problematic inter-component dependencies, architectural erosion, or the continued use of architecturally significant obsolete technologies~\cite{li2022identifying}. This distinction is practically important because ATD often affects multiple components and typically involves more costly and cross-cutting repayment decisions than localized forms of debt \cite{carrillo2018ripple}.

Later studies extended SATD detection beyond source code comments to issue trackers. Dai and Kruchten~\cite{dai2017detecting} showed that machine learning can identify SATD in issue reports, while Li et al.~\cite{li2022identifying} and Li et al.~\cite{li2023automatic} further applied deep learning and found that issue trackers contained the highest prevalence of ATD. Recently, Shivashankar et al.~\cite{shivashankar2025beacon} proposed a transformer-based approach that classifies technical debt types, including architecture debt. However, their main focus, as in prior work, remained broad debt type classification rather than ATD-specific detection.

Taken together, existing work on ATD can be broadly grouped into three streams: structural detection through architectural smells and dependency analysis, empirical studies on ATD impact and evolution, and broad technical debt classification studies in which architecture debt appears as one category among several debt types. However, these streams do not yet provide a dedicated pipeline for identifying self{-}admitted ATD directly from raw Jira issue reports while also reducing annotation effort and supporting explainable model decisions.

Overall, prior work has improved our understanding of ATD and its consequences, but it generally assumes that ATD items are already available for analysis. Our work complements this body of research by addressing an earlier bottleneck, namely how to efficiently identify and label ATD instances from issue tracking systems at scale. Motivated by this perspective, our study focuses on ATD as expressed in Jira issue reports. We focus on Jira-based open-source projects because they provide rich and structured issue descriptions, transparent development histories, and strong reproducibility~\cite{yli2016software,dabbish2012leveraging}. This setting is particularly suitable for studying ATD acknowledged in Jira issue reports, as prior work suggests that issue trackers provide strong signals of architecture-related debt~\cite{li2023automatic}.

\subsection{ Active Learning in SE}
Active learning (AL) is a learning paradigm in which a model iteratively selects the most informative unlabeled instances for annotation, aiming to improve performance while reducing labeling effort~\cite{settles2009active}. In SE, this paradigm is especially attractive because labeled data are often expensive to obtain, and manual analysis frequently requires expert judgment. Early work in software behavior classification showed that AL can achieve classifier quality comparable to batch learning while using substantially less labeling effort, highlighting its value when human analysis is costly~\cite{bowring2004active}. More recent work has also positioned AL more broadly as a mechanism for improving SE workflows by leveraging limited supervision and exploiting the structure of software-related data~\cite{cambronero2019active}. This makes AL particularly relevant in tasks where labels are scarce, expensive, or must be acquired iteratively.

Among the main AL settings, pool-based active learning is especially suitable for text classification because it assumes a small labeled set together with a large pool of unlabeled instances. In this setting, the learner repeatedly selects informative candidates from the unlabeled pool, queries an oracle for labels, updates the classifier, and continues the process. This iterative workflow has been used in SE to expand the scope of behavior models efficiently~\cite{bowring2004active} and, more recently, to reduce the number of bug reports that must be manually labeled while maintaining strong predictive performance~\cite{wu2021improving}. In this study, we follow this general setting to construct an ATD classifier for Jira issues while minimizing the number of labeled instances required. To enable controlled and reproducible experiments, we simulate the annotation process using fully labeled datasets, where the training partition serves as the unlabeled pool from which instances are iteratively selected.

A central motivation for using AL in SE is the high cost of labeling. In SATD identification, for example, prior work has shown that generating reliable labels is labor-intensive and that even semi-automated pipelines may still require substantial manual inspection~\cite{tu2022debtfree}. Similarly, in high-impact bug report prediction, AL has been used to iteratively select uncertain reports for expert review, substantially reducing the number of instances that require labeling while preserving predictive quality~\cite{wu2021improving}. Taken together, these studies suggest that AL is particularly useful in SE tasks where the annotation budget is limited, but label quality remains critical.

A key component of AL is the query strategy used to decide which instances should be labeled next. We consider several widely used strategies representing uncertainty-based and diversity-based sampling:

\begin{itemize}
    \item \textbf{Prediction Entropy}~\cite{holub2008entropy}, which selects instances with the highest predictive uncertainty.
    \item \textbf{Least Confidence}~\cite{li2006confidence}, which selects instances for which the model assigns the lowest confidence to its top predicted class.
    \item \textbf{Breaking Ties}~\cite{luo2005active}, which selects instances for which the two most probable classes have the smallest margin.
    \item \textbf{Embedding K-Means}~\cite{yuan2020cold}, which promotes diversity by selecting instances near cluster centroids in the embedding space.
    \item \textbf{Contrastive Active Learning}~\cite{margatina-etal-2021-active}, which prioritizes instances that are most informative relative to labeled examples or competing predictions.
    \item \textbf{Random sampling}, which serves as a baseline.
\end{itemize}


Although AL has already been explored in several SE contexts, its application to ATD detection introduces two additional challenges. First, ATD is an underrepresented form of technical debt, and high-quality labels require architectural expertise, making annotation substantially more expensive than in more common SE classification tasks. Second, ATD signals in issue reports are often implicit, context-dependent, and lexically diverse, thereby increasing uncertainty in manual labeling and making sample selection especially important. Therefore, in this study, AL is used not simply as a generic classification optimization technique but as a mechanism to make the construction of ATD training data more feasible and scalable.

\subsection{Explainable AI in SE}
Explainable AI (XAI) aims to make the decisions of machine learning models more understandable to humans~\cite{dwivedi2023explainable}. In SE, this is especially important because AI-based tools are increasingly used across many SE tasks, yet their practical adoption is often hindered by the limited transparency of complex models~\cite{tantithamthavorn2021explainable}. Recent evidence shows that XAI has been applied across a broad range of SE problems, but challenges remain regarding evaluation, stakeholder needs, and practical usefulness~\cite{cao2025systematic}.

Among model-agnostic XAI methods, SHAP~\cite{lundberg2017unified} and LIME~\cite{ribeiro2016should} are widely used for local explanations. SHAP estimates the contribution of each feature using Shapley values from cooperative game theory, thereby providing additive, theoretically grounded explanations~\cite{shapley1953value}. LIME explains individual predictions by perturbing the input and fitting an interpretable surrogate model around the prediction of interest~\cite{ribeiro2016should}. In SE, these methods have been used to explain defect prediction, issue prediction, and technical debt identification models, showing that local explanations can help users inspect model behavior and understand influential features or textual cues~\cite{ribeiro2016should,lundberg2017unified,esteves2020understanding,tsoukalas2024local,schulte2024studying}.

Prior studies also suggest that XAI can improve trust and support model validation in SE contexts. For example, work on issue prediction found that both LIME and SHAP can provide reasonable explanations, whereas SHAP may produce less ambiguous and more contextual explanations because it can capture sentence fragments rather than isolated tokens. Similarly, studies on technical debt and defect prediction show that combining local and global explanations can help practitioners inspect both overall model behavior and case-specific decisions~\cite{schulte2024studying,tsoukalas2024local,gezici2025explainable}.

In this study, we use SHAP and LIME to interpret ATD classification results from Jira issues. This setting differs from most prior SE applications because our input is textual issue data and the target concept is architecture technical debt, which is subtle, context-dependent, and expert-driven. By highlighting the words or phrases that influence each prediction, LIME and SHAP help experts inspect whether the classifier relies on meaningful architectural signals. We further assess these explanations through expert evaluation, following recent calls for XAI studies in SE to consider explanation quality and stakeholder usefulness more explicitly \cite{cao2025systematic,schulte2024studying,avgeriou2025manifesto}.

\section{Study Design}
\label{sectionRM}

\subsection{Research Questions}
The goal of this study, formulated according to the Goal-Question-Metric (GQM) \cite{van2002goal} template is to \enquote{\textit{\textbf{analyze} Jira issues 
\textbf{in order to} reduce annotation effort in architecture technical debt detection \textbf{with respect to} (i) the effectiveness of keyword-based filtering candidate issues, (ii) the efficiency and performance of active learning strategies, and (iii) the explainability of model decisions using XAI techniques \textbf{from the point of view of} SE experts \textbf{in the context of} open-source software projects}}.

Based on this goal, we derived the following research questions (RQs):

\textbf{RQ1:} \textit{How effective are keyword extraction techniques in identifying candidate ATD issues from Jira issue trackers}\\
\textit{\textbf{Rationale}:} Since ATD is relatively rare and difficult to identify directly at scale, we first investigate whether keywords derived from expert-validated ATD issues can be used to retrieve relevant candidate issues from large Jira datasets. This stage evaluates whether keyword-based filtering can serve as a practical triage mechanism prior to manual validation and model training.\\

\textbf{RQ2:} \textit{To what extent can active learning improve ATD classification performance while reducing annotation effort?}\\
\textit{\textbf{Rationale}:} ATD detection is a low-resource classification problem in which labeled data is costly to obtain because annotation requires architectural judgment and many issues contain implicit or borderline signals. Therefore, we investigate whether active learning can reduce labeling effort while still achieving strong classification performance.\\

\textbf{RQ3:} \textit{To what extent do LIME and SHAP explanations support the interpretability of ATD classification results for software engineering experts?}\\
\textit{\textbf{Rationale}:} Because transformer-based classifiers behave as black-box models, their predictions may be difficult for experts to understand and trust in ATD identification tasks. This question evaluates whether LIME and SHAP provide explanations that are sufficiently clear and useful to support expert judgment.\\


\subsection{Research Methodology}
\label{subsection:research_method}
\begin{figure*}[htp] 
    \centerline{\includegraphics[trim=0.1cm 0.6cm 0.1cm 0.9cm, clip, width=1.0\textwidth]{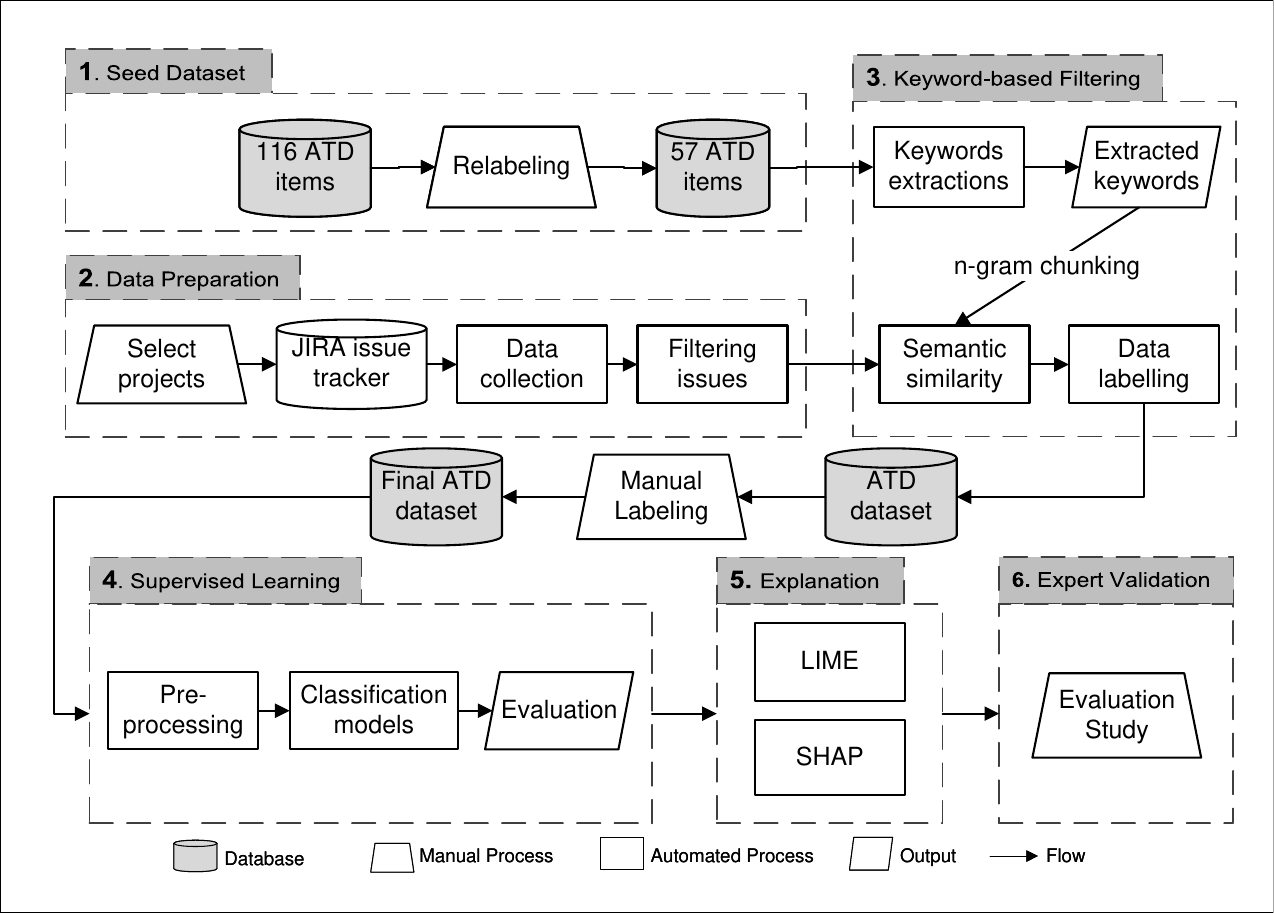}}
    \caption{Overview of the proposed approach.}
    \label{fig_methodology}
 \end{figure*}

This study was designed as a sequential pipeline to identify ATD from Jira issue reports. As shown in Figure~\ref{fig_methodology}, the pipeline consists of six stages: (1) seed dataset construction and annotation scheme definition, (2) data preparation, (3) keyword-based filtering, (4) supervised learning with active learning, (5) explainability, and (6) expert validation. Each stage uses the output of the previous stage.

The three research questions correspond to the main stages of the pipeline. RQ1 concerns the keyword-based filtering stage, RQ2 the supervised learning stage with active learning, and RQ3 the explainability and expert validation stages. In the following subsections, we describe each stage and introduce the corresponding evaluation procedures in context.

\paragraph{\textbf{1. Seed Dataset and Annotation Scheme.}
To guide the annotation process, we adopted the classification framework of Li et al. \cite{li2020identification}, which builds on the original framework proposed by Alves et al. \cite{alves2014towards} by defining several debt types and indicators. In this study, we focused on the two indicators related to architectural technical debt, namely violation of modularity (VioMod) and using obsolete technology (ObsTech). \textbf{VioMod} refers to situations in which shortcuts introduce undesired interdependencies among multiple modules that should remain independent. \textbf{ObsTech} refers to cases where an architecturally significant technology has become obsolete.}

During manual labeling, these indicators were used to assess whether a Jira issue contained clear architectural debt signals, only borderline architectural concerns, or no architectural debt evidence. Based on this conceptual grounding, we distinguish between two levels of architectural debt expression in issue tracking systems: \textit{True{-}ATD} and \textit{Weak{-}ATD}.


\textbf{True{-}ATD} refers to issue reports that explicitly and unambiguously describe architectural concerns, trade-offs, or structural design decisions that may compromise long-term maintainability, scalability, or performance. These issues are typically labeled \enquote{ATD} by both annotators without ambiguity and often contain clear architectural terminology, references to components or layers, or discussions about architectural rationale. When disagreements occurred between annotators (e.g., \enquote{ATD} vs. \enquote{Non{-}ATD}), a third expert was consulted, and majority voting determined the final classification.

\textbf{Weak{-}ATD} refers to issue reports that show possible architectural concerns but lack sufficient context or clarity for definitive classification. These include cases in which one annotator classified the issue as \enquote{ATD}, while the other assigned a \enquote{Maybe} label, indicating the presence of architectural concerns but insufficient clarity for a definitive classification. While such issues may hint at design limitations, modularity concerns, or dependency restructuring, they do so in vague language and without detailing the architectural impact. Weak{-}ATD denotes borderline cases that fall short of the criteria for True{-}ATD but remain informative for training or analysis under a more lenient interpretation of architectural debt.

To make these definitions more concrete, we next present a representative example of True{-}ATD and contrast it with an example of code debt, a different SATD type. This comparison helps clarify how ATD differs from other forms of SATD.

\begin{figure*}[h] 
    \centerline{\includegraphics[trim=0.1cm 0.1cm 0.1cm 0.1cm, clip, width=0.95\textwidth]{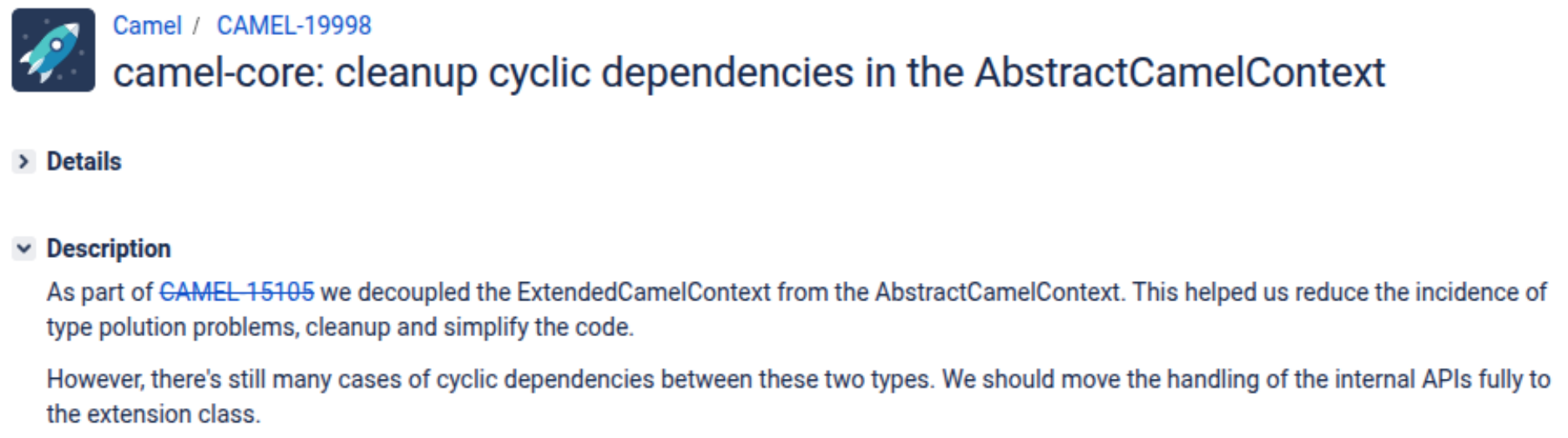}}
    \caption{Example of an issue labeled as True{-}ATD from CAMEL-19998.\protect\footnotemark}
    \label{fig_True_ATD}
\end{figure*}
\footnotetext{\url{https://issues.apache.org/jira/browse/CAMEL-19998}}

As shown in Figure~\ref{fig_True_ATD}, the issue is labeled as True{-}ATD because it explicitly describes cyclic dependencies and the need to relocate internal API handling, which together indicate a structural modularity problem rather than a local code{-}level concern. This example also clarifies how ATD differs from broader SATD types. While SATD may include any self{-}admitted technical compromise, ATD in this study is reserved for cases that describe architectural consequences, such as problematic dependencies, modularity violations, or component-level restructuring.

\begin{figure*}[h!]
    \centerline{\includegraphics[trim=0.4cm 0.1cm 0.1cm 0.1cm, clip, width=0.95\textwidth]{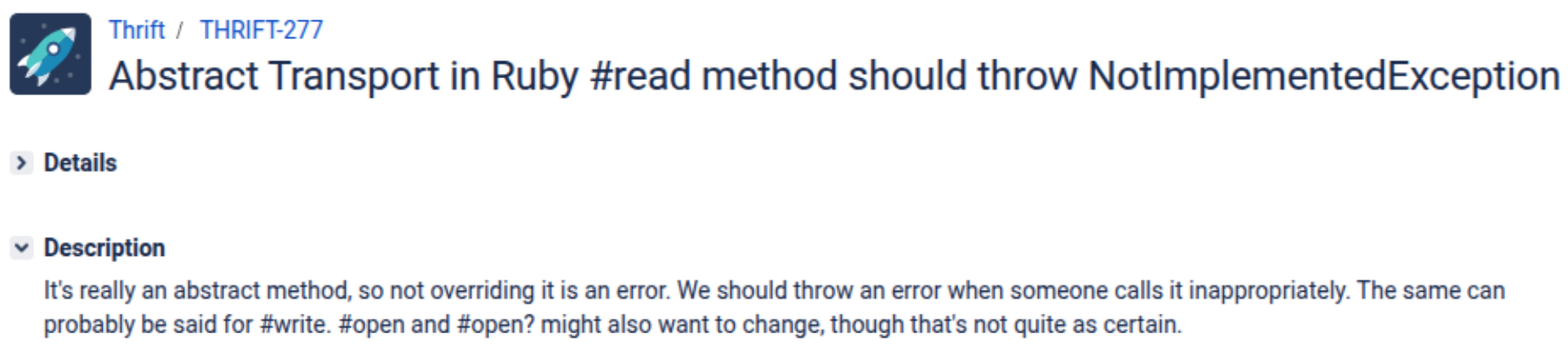}}
    \caption{Example of an issue labeled as code debt (but neither True- nor Weak-ATD) from THRIFT-277.\protect\footnotemark}
    \label{fig_code_debt}
\end{figure*}
\footnotetext{\url{https://issues.apache.org/jira/browse/THRIFT-277}}

To further clarify how ATD differs from other SATD types, Figure~\ref{fig_code_debt} presents an example of code debt from THRIFT-277, labeled as code debt by Li et al.~\cite{li2022identifying}. The issue concerns the implementation of an abstract method and the need to throw an exception when it is called inappropriately. This reflects a localized implementation-level deficiency rather than an architectural concern affecting system structure. By contrasting Figures~\ref{fig_True_ATD} and~\ref{fig_code_debt}, we clarify that SATD is the broader concept of self{-}admitted technical debt in textual artifacts, whereas ATD refers specifically to self{-}admitted debt with architectural implications.


In preparation for the keyword extraction stage, we relabeled an initial dataset of 116 ATD items compiled by Li et al. \cite{li2023automatic}, refining it to 57 items agreed upon by all authors (the \textit{Relabeling} activity in Figure~\ref{fig_methodology}). This relabeling step was necessary to improve the overall reliability of the dataset and to reduce potential false positives carried over from earlier annotations. Each author independently labeled all items, and only those unanimously labeled as ATD by all three authors were retained as seed items. Because these items were used as the seed dataset for keyword extraction, we adopted this conservative criterion to ensure that the seeds represented clear and reliable examples of architectural technical debt. These curated ATD items, therefore, served as a robust starting point for identifying the characteristics and terminology commonly associated with architecture-related technical debt.

\paragraph{\textbf{2. Data Preparation.}}
After constructing the seed dataset, we prepared a large corpus of Jira issues from open-source projects for large-scale ATD identification. To construct the dataset, we selected new projects from a curated list of Java frameworks, libraries, and software.\footnote{\url{https://java.libhunt.com/}} Since not all projects in this list use Jira for issue tracking, we filtered the candidates and retained only those that use Jira and satisfy our scale and activity criteria.

Project selection was guided by availability, diversity in application domains, and sustained development activity to ensure a sufficient number of issues. We included only projects with more than 5,000 reported issues, following the recommendation of Li et al. \cite{li2020identification} to ensure sufficient complexity. Many of the selected projects are hosted by the Apache Software Foundation, which enforces standardized development practices, transparent issue tracking, and consistent versioning, making them well-suited for rigorous and reproducible research. Issue summaries and descriptions were collected using Python and the Jira API.

The final dataset comprised ten open-source projects that use Jira as their issue-tracking system. This number aligns with previous SATD studies, which typically analyze between four and ten Java projects \cite{potdar2014exploratory,li2020identification,da2017using}. Table~\ref{tab:projects} summarizes the selected projects, including their application domain, total source lines of code (SLOC), and the number of reported and resolved issues. We analyzed the most recent versions as of January 7, 2025, and measured SLOC using the SCC tool.\footnote{\url{https://github.com/boyter/scc}}

Our focus on Java projects is motivated by three considerations. First, Java is one of the most widely used programming languages in open-source systems, particularly in enterprise-scale applications that often involve complex architectures prone to technical debt. Second, prior studies on SATD and ATD have also mostly used Java projects \cite{potdar2014exploratory,li2023automatic,sheikhaei2024empirical}, which supports comparability with existing work. Third, the availability of large, active, and well{-}documented Java projects using Jira makes them suitable for systematic data collection and analysis.

\begin{table}[htp]
  \caption{Projects used in this study.} 
  \scriptsize
  \label{tab:projects}
  \begin{tabular}{m{0.1cm}>{}m{2.23cm}>{}m{3.3cm}>{}m{1.2cm}>{}m{1.1cm}>{}m{1.1cm}}
    \toprule
    No & Project & Domain & SLOC & \#reported issues & \#resolved issues\\
    \midrule
    1 & Apache Camel & Integration framework & 1,800k & 22,068 & 16,478\\
    2 & Apache Spark & Analytics engine & 1,442k & 55,155 & 42,348\\    
    3 & Apache Kafka & Stream-processing software & 1,012k & 19,283 & 11,403\\
    4 & Apache ActiveMQ & Message broker & 423k & 9,712 & 5,512\\
    5 & Apache Cassandra & Database & 798k & 20,648 & 16,967\\
    6 & Apache Drill & Query engine & 890k & 8,524 & 3,571\\
    7 & Apache Geode & Data management & 1,350k & 10,457 & 1,125\\
    8 & Apache Lucene & Search engine library & 882k & 10,681 & 2,438\\
    9 & Apache Netbeans & IDE & 5,400k & 6,519 & 272\\
    10 & Apache Solr & Load balancer & 705k & 17,762 & 3,847\\
  \bottomrule
\end{tabular}
\end{table}

Prior studies, such as Li et al.~\cite{li2022identifying}, analyzed Jira issues at the sentence level by labeling individual sentences as technical debt. However, this approach is less suitable for ATD detection because issue summaries are often vague~\cite{janak2009issue}, making it difficult to infer architectural context from fragmented text. Following Diamantopoulos et al.~\cite{diamantopoulos2023semantically}, we combined each issue's summary and description into a single text to preserve context and better capture the issue as a whole.

In the filtering process, we focused exclusively on issues marked as resolved. This provides a complete snapshot of the issue lifecycle, including how it was ultimately addressed or mitigated. By focusing on resolved issues, we also minimized noise generated by incomplete or ongoing work, thereby improving the reliability of the resulting dataset.

\paragraph{\textbf{3. Keyword-Based Filtering.}}
Using the refined seed ATD dataset, we next extracted ATD-related keywords and applied them to identify candidate ATD issues in the new Jira repositories. A central activity in this stage was keyword extraction, which aimed to identify terms and phrases indicative of ATD. We applied three methods: Term Frequency{-}Inverse Document Frequency (TF{-}IDF) \cite{ramos2003using}, KeyBERT \cite{grootendorst2020keybert}, and Class-Specific KeyBERT (CS KeyBERT) \cite{meisenbacher2024improved}.

These methods were selected because they represent different paradigms for keyword identification. TF{-}IDF is a statistical method that prioritizes terms that are frequent within a document but rare across the corpus. In contrast, KeyBERT uses transformer{-}based embeddings, such as Sentence-BERT \cite{reimers2019sentence}, to identify keywords that are semantically similar to the overall content of the text. CS KeyBERT extends KeyBERT with a guided mechanism that incorporates seed keywords.

In our study, we employed predefined seed keywords, such as \enquote{\textit{move}}, \enquote{\textit{refactor}}, \enquote{\textit{remove}}, \enquote{\textit{dependency}}, \enquote{\textit{couple}}, and \enquote{\textit{update}}, to guide keyword extraction toward potentially ATD-related vocabulary. These seed keywords were not treated as architectural indicators in isolation. Rather, they were selected because they frequently occurred in the expert-validated seed ATD issues derived from the ATD items of Li et al.~\cite{li2023automatic} and helped steer CS KeyBERT toward phrases that describe architectural actions and consequences. Their architectural relevance therefore emerges primarily through the bi{-}gram and tri{-}gram expressions and semantic contexts they help surface, rather than through the isolated uni{-}grams themselves.

The three methods were used to extract uni{-}gram, bi{-}gram, and tri{-}gram tokens that strongly indicate ATD. To refine the extracted keywords and align them with architecture-related concerns, we applied text mining techniques, including tokenization, stop-word removal, stemming or lemmatization, and part-of-speech tagging using spaCy \cite{honnibal2020spacy}. 

After keyword extraction, we matched the extracted keywords against Jira issue texts using n{-}gram chunking and semantic similarity. Specifically, the extracted keywords were aligned with n{-}grams in Jira issue reports using BERT embeddings, and cosine similarity scores were computed between keywords and issue-description fragments. Given two vector representations $\mathbf{A}$ and $\mathbf{B}$ of a keyword and an n{-}gram phrase, respectively, cosine similarity is defined as:

\begin{equation}
\text{Cosine Similarity} = \cos(\theta) = \frac{\mathbf{A} \cdot \mathbf{B}}{\|\mathbf{A}\| \|\mathbf{B}\|} = \frac{\sum_{i=1}^{n} A_i B_i}{\sqrt{\sum_{i=1}^{n} A_i^2} \sqrt{\sum_{i=1}^{n} B_i^2}}
\end{equation}

where $\mathbf{A} \cdot \mathbf{B}$ is the dot product of vectors $\mathbf{A}$ and $\mathbf{B}$, $\|\mathbf{A}\|$ and $\|\mathbf{B}\|$ are their Euclidean norms, and $\cos(\theta)$ measures the angle-based similarity between the two vectors.

We applied a sliding-window approach to compare the extracted keywords with all possible n{-}grams in the issue descriptions. If an issue contained at least one n{-}gram with a similarity score above 0.9, it was automatically labeled as an ATD candidate.

\begin{figure}[htp] 
    \centerline{\includegraphics[trim=0.2cm 0.9cm 0.5cm 0.6cm, clip, width=0.8\textwidth]{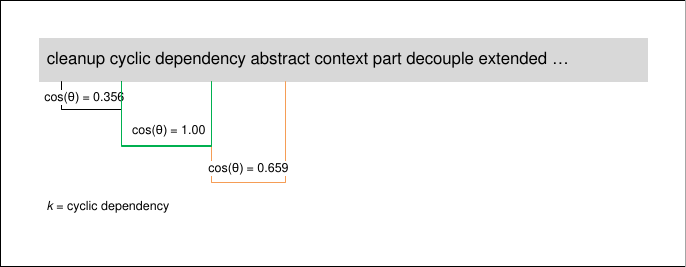}}
    \caption{Example of bi{-}gram chunking and similarity scoring for CAMEL-19998. The phrase \enquote{cyclic dependency} is identified as the most semantically similar chunk to ATD-related keywords, based on cosine similarity scores.}
    \label{fig_chunking}
\end{figure}
\vspace{-1pt}

Figure~\ref{fig_chunking} illustrates this process with an example from issue CAMEL-19998. A sliding window is used to generate n{-}gram candidates from the issue description, which are then compared to the extracted ATD-related keywords. In this example, the bi{-}gram \enquote{cyclic dependency} obtains the highest cosine similarity score, indicating a strong association with architectural technical debt.

This automated matching process was applied across the selected open-source projects, resulting in a large-scale keyword-filtered dataset of more than 103,000 labeled issues consisting of both ATD and Non{-}ATD classes.

To answer RQ1, we evaluated the effectiveness of keyword extraction by manually reviewing statistically representative samples from both issues labeled as ATD candidates and issues labeled as Non{-}ATD by the keyword-based approaches. This evaluation allowed us to assess the usefulness of the extracted keywords in retrieving relevant ATD issues while also estimating the extent of false-negative cases.

To validate the effectiveness of the keyword set used for ATD identification, we conducted a qualitative evaluation of statistically representative samples of issues labeled as ATD or Non{-}ATD by the keyword-based approaches. Two authors independently reviewed 2,161 ATD samples and 1,152 Non{-}ATD samples, where each instance corresponds to a single labeled Jira issue.

The required sample sizes were calculated using a 95\% confidence level and a 5\% margin of error according to the following formula:

\begin{equation}
\text{Sample size} =
\frac{
\frac{z^{2} \times p(1-p)}{e^{2}}
}{
1 + \left(
\frac{z^{2} \times p(1-p)}{e^{2}N}
\right)
}
\end{equation}

In this formula, \(z\) represents the z-score for the selected confidence level, \(p\) the estimated population proportion, \(e\) the margin of error, and \(N\) the population size. Based on this calculation, the samples were randomly selected while preserving stratified representation across the keyword extraction techniques, including manually defined seed keywords and n{-}gram-based approaches. When disagreements arose between the two annotators, a third author was consulted, and the final label was assigned by majority vote.

\label{item:supervised-approach} \paragraph{\textbf{4. Supervised Learning with Active Learning.}}
After constructing the labeled dataset, we used it to train supervised classification models for ATD detection. Since supervised learning requires annotated examples, the labeled issue dataset provides the basis for learning textual patterns that distinguish ATD from Non{-}ATD issues.

To reduce labeling effort, we employed active learning with a BERT-based model that iteratively selects the most informative samples for annotation. We used BERT as the base learner in the active learning loop because it provided a strong and stable transformer baseline while keeping iterative retraining computationally feasible. This choice is also supported by prior evidence from a systematic review, which found that BERT is among the most effective transformer-based models for technical debt detection tasks~\cite{sutoyo2026self}. Since each active learning strategy requires repeated model updating across multiple annotation rounds, evaluating the full set of query strategies with every transformer architecture would substantially increase the computational cost of the study. Therefore, we used BERT for the active learning comparison and included RoBERTa and DeBERTa{-}v3 in the supervised benchmark to position the results against stronger Transformer baselines.

Several query strategies were explored to determine which approach achieves the best performance with minimal labeled data. Before training the models, we applied minimal preprocessing to clean and standardize the textual data, including lowercasing and removing irrelevant characters such as code snippets and formatting tags.

After preprocessing, we trained and evaluated a comprehensive suite of classifiers. This included BERT with active learning using various query strategies and Random Sampling as a baseline. In addition, to provide a rigorous benchmark, we included machine learning and deep learning algorithms widely adopted in SATD detection research.

Following Li et al. \cite{li2022identifying}, we included Support Vector Machine (SVM), Na\"ive Bayes (NB), k-Nearest Neighbors (kNN), Logistic Regression (LR), Random Forest (RF), and TextCNN. We also considered transformer-based approaches following Sharma et al. \cite{sharma2022self}, including ALBERT and RoBERTa, and further compared BERT, RoBERTa, and DeBERTa{-}v3 following Skryseth et al. \cite{skryseth2023technical}.

To answer RQ2, we compared the performance of active learning configurations and supervised baselines on the ATD classification task. Because the class distribution is imbalanced, we used precision, recall, and F1 score as primary evaluation metrics to assess whether active learning can improve classification performance with fewer labeled instances.

\paragraph{\textbf{5. Explainability.}} To address RQ3, we applied XAI techniques, namely LIME and SHAP, to interpret the predictions made by the BERT-based ATD classifier. Because transformer models are inherently opaque, these techniques help reveal which tokens or textual features most strongly influenced the model's decision to classify a Jira issue as ATD or Non{-}ATD.

LIME provides local explanations by learning an interpretable surrogate model around each prediction \cite{ribeiro2016model}. SHAP, grounded in game theory, assigns contribution values to features based on their marginal impact on model predictions \cite{lundberg2017unified}. Together, these methods provide complementary local explanations that reveal which parts of the issue description contributed most to the prediction.

Both LIME and SHAP were selected because they are widely used, model{-}agnostic XAI techniques that have demonstrated usefulness in increasing model interpretability and trustworthiness \cite{salih2025perspective}. LIME is particularly effective for intuitive instance-level explanations, whereas SHAP is valuable for detailed contribution analysis. Using both methods allows us to investigate explainability in ATD detection from complementary perspectives.

\paragraph{\textbf{6. Expert Validation.}} 
The final stage of the methodology evaluates the interpretability and usefulness of the generated explanations through expert judgment. Following Stol and Fitzgerald's ABC framework \cite{stol2018abc}, we conducted an evaluation study, a form of judgment study in which domain experts assess a research artifact in a structured way.

In this study, SE experts were presented with model predictions and explanation outputs generated by LIME and SHAP. They were then asked to evaluate each explanation across dimensions such as clarity, relevance to architectural concerns, actionability, and support for trust in the model's decision. This evaluation provides insight into how well current XAI techniques support ATD-related decision making. To answer RQ3, we evaluated explanation quality through an expert judgment study using a structured questionnaire. Experts assessed LIME and SHAP explanations across eight dimensions derived from prior XAI work, namely trustworthiness, informativeness, interactivity, fairness, confidence, accessibility, causality, and transferability.

To recruit experts, we targeted SE researchers at the rank of associate professor or above, with at least five years of experience in technical debt, SATD, and software architecture. Candidates were identified from prominent publications in the field and through recommendations from academic advisors and established researchers. Invitations were distributed via email and personal contacts within the SE research community.

We invited 17 experts to participate, and 10 experts contributed to the evaluation study. Each expert completed an online survey in which they reviewed predictions and corresponding explanations, then rated each explanation across key dimensions informed by the XAI goals identified by Al{-}Ansari et al. \cite{al2024user}, namely trustworthiness, informativeness, interactivity, fairness, confidence, accessibility, causality, and transferability. Table~\ref{tab:xai-goal} summarizes these goals and their definitions, adapted from \cite{arrieta2020explainable,al2024user}. The complete questionnaire is provided in the replication package.

\begin{table}[ht]
  \caption{Summary of XAI goals with definitions adapted from \cite{arrieta2020explainable,al2024user}.} 
  \label{tab:xai-goal}
  \scriptsize
  \begin{tabular}{m{1.9cm}>{}m{9.1cm}}
    \toprule
    Goal & Definition \\
    \midrule
    Trustworthiness & is the confidence that a model will act as intended when facing a given problem \\
    Informativeness & is about the model providing sufficient information to support human decision{-}making \\
    Interactivity & refers to the model's ability to allow user interaction, letting users tweak or query the model to gain better understanding \\
    Fairness & is the capacity of the model to ensure and guarantee unbiased, equitable decision-making \\
    Confidence & An explainable model should provide information about its confidence or the certainty of its predictions \\
    Accessibility & is about making models understandable and usable for end users, including those who are non{-}technical \\
    Causality & Explainable models might help in identifying potential causal relationships, which can be tested further \\
    Transferability & refers to the model's ability to be applied or reused in different contexts or problems \\
    \bottomrule
  \end{tabular}
\end{table}

\section{Results}
\label{sectionResults}

\subsection{RQ1. Effectiveness of the Extracted Keywords} 
\label{subsectionRQ1}

To answer RQ1, we evaluated whether keywords extracted from the refined seed ATD dataset can be reused to identify candidate ATD issues in new Jira repositories. We compared three keyword extraction methods, namely TF-IDF, KeyBERT, and CS KeyBERT, and then validated the resulting labels through manual review of statistically representative samples from both the predicted ATD and predicted Non{-}ATD sets.

\subsubsection{Keyword Extraction Results}

Tables~\ref{tab:tf-idf}--\ref{tab:class-specific} present the top-15 extracted keywords for each n-gram level produced by TF-IDF, KeyBERT, and CS KeyBERT, respectively. These tables are intended to illustrate the most representative ATD{-}related terms identified by each method, while the complete keyword sets are available in the replication package. Across the methods, many extracted keywords reflect recurring architectural concerns, such as dependency restructuring, modularity violations, refactoring, component movement, and technology replacement. This indicates that the refined seed dataset contains sufficiently consistent architectural vocabulary to support keyword reuse across projects.

\begin{table}[htp]
  \caption{Top{-}15 keywords extracted using TF-IDF for each n{-}gram level.}
  \label{tab:tf-idf}
  \scriptsize
  \begin{tabular}{m{1.5cm}>{}m{3.1cm}>{}m{4.1cm}}
    \toprule
    uni{-}gram & bi{-}gram & tri{-}gram\\
    \midrule
     use & dangerous upgrade & break multiple file\\
     test & test depend & build place file\\
     package & upgrade modern & place file instead\\
     upgrade & place file & code use test\\
     file & break multiple & file wrong place\\
     version & multiple file & package place file\\
     need & rest api & place file wrong\\
     like & file wrong & test let relocate\\
     class & package place & use test let\\
     dependency & wrong place & dependency longer need\\
     remove & need make & like upgrade lose\\
     place & code use & lose stuff dependency\\
     old & let relocate & stuff dependency need\\
     add & test let & upgrade lose stuff\\
     new & use test & backup test fully\\
  \bottomrule
\end{tabular}
\end{table}

\begin{table}[htp]
  \caption{Top{-}15 keywords extracted using KeyBERT for each n{-}gram level.}
  \label{tab:keybert}
  \scriptsize
  \begin{tabular}{m{1.5cm}>{}m{3.1cm}>{}m{4.1cm}}
    \toprule
    uni{-}gram & bi{-}gram & tri{-}gram\\
    \midrule
     move & use late & move client class\\
     could & get rid & class client package\\
     test & rest api & improvement class loading\\
     version & year old & late library improvement\\
     package & move client & library improvement class\\
     dependency & client package & use late library\\
     upgrade & class client & upgrade use late\\
     class & next release & nice could upgrade\\
     use & build mvc & could upgrade use\\
     also & place file & rest client release\\
     need & library improvement & current transport client\\
     like & late library & high level rest\\     
     api & could upgrade & client instead new\\
     build & class loading & support basic authentication\\
     library & upgrade use & rest api client\\
  \bottomrule
\end{tabular}
\end{table}

\begin{table}[htp]
  \caption{Top{-}15 keywords extracted using CS KeyBERT for each n{-}gram level.}
  \label{tab:class-specific}
  \scriptsize
  \begin{tabular}{m{1.5cm}>{}m{3.1cm}>{}m{4.1cm}}
    \toprule
    uni{-}gram & bi{-}gram & tri{-}gram\\
    \midrule    
    move & specific dependency & independent indirect dependency\\
    refactor & refactor code & need late version\\
    remove & dependency need & outdated include exception\\
    dependency & stuff dependency & dependency version need\\
    couple & dependency specification & specific dependency version\\
    update & dependency good & dependency good place\\
    improve & dependency version & create cyclic dependency\\
    relocate & dependency exist & old package depend\\
    transfer & upgrade dependency & indirect dependency exist\\
    migrate & class refactor & package outdated include\\
    problem & remove dependency & release year old\\
    increase & try depend & turn package outdated\\
    extend & cyclic dependency & look dependency package\\
    depend & need update & hard work update\\
    change & version need & class refactor code\\
    \bottomrule
\end{tabular}
\end{table}

Among the three methods, CS KeyBERT appears to generate the most architecture-focused vocabulary, including terms such as \enquote{remove dependency}, \enquote{cyclic dependency}, and \enquote{independent indirect dependency}. In contrast, TF-IDF yields a broader set of frequent lexical patterns, while KeyBERT captures semantically meaningful but more general expressions. These differences suggest that the three methods provide complementary perspectives on ATD-related language rather than identical keyword sets.

\subsubsection{Manual Validation Results}

Table~\ref{tab:statistics} summarizes the number of issues labeled as ATD candidates by each keyword extraction method under the semantic similarity threshold of 0.9. TF-IDF produced the largest number of candidate issues, followed by KeyBERT and CS KeyBERT. This indicates that keyword extraction can scale the identification process from a small refined seed set to a much larger candidate pool.

\begin{table}[htp]
  \caption{Number of ATD candidate issues labeled by each method with a threshold $> 0.9$. The size of the samples extracted from each method for manual annotation is shown in square brackets [].} 
  \label{tab:statistics}
  \scriptsize
  \begin{tabular}{m{2.5cm}>{}m{2.5cm}>{}m{2.5cm}>{}m{2.5cm}}
    \toprule
    Method & TF-IDF & KeyBERT & CS KeyBERT \\
    \midrule
    uni{-}gram & 24,823 [379] & 18,400 [377] & 12,195 [373]\\
    bi{-}gram & 1,005 [279] & 532 [224] & 963 [275]\\
    tri{-}gram & 12 [12] & 13 [13] & 28 [27]\\
    \midrule
    Total & 25,840 [670] & 18,945 [614] & 13,186 [675]\\
  \bottomrule
\end{tabular}
\end{table}

To assess the reliability of the keyword-based labeling process, we manually reviewed 2,161 ATD{-}labeled instances sampled from the candidate pool. The annotation required substantial effort because each issue included both a summary and a description. Across the two annotators, the ATD-side validation required more than 72 hours of combined effort. The resulting Cohen's kappa coefficient was +0.70, indicating substantial agreement \cite{landis1977measurement}.

Table~\ref{tab:dataset} presents the validated composition of True{-}ATD and Weak{-}ATD identified by each keyword-based method. When considering only True{-}ATD, the three methods correctly identified between 127 and 222 cases, corresponding to 21\%--33\% of the manually reviewed ATD-labeled samples. When Weak{-}ATD is included, the coverage increases substantially, ranging from 25\% to 70\%. This suggests that keyword-based methods are better at capturing broader architectural signals than strictly explicit ATD cases.

\begin{table}[htb]
  \caption{Validated composition of True{-}ATD. The number of Weak{-}ATD items is shown in square brackets.} 
  \label{tab:dataset}
  \scriptsize
  \begin{tabular}{m{3cm}>{}m{2.3cm}>{}m{2.3cm}>{}m{2.3cm}}
    \toprule
    Method & TF-IDF & KeyBERT & CS KeyBERT \\
    \midrule
    uni{-}gram & 84 [118]  & 78 [14] & 106 [25]  \\ 
    bi{-}gram & 72 [186]  & 47 [13] & 109 [42]  \\
    tri{-}gram &  4 [6]  & 2 [1] & 7 [6] \\
    \midrule
    Total & 160 [310] & 127 [28] & 222 [73]\\
    \midrule
    $\Sigma$ True{-}ATD & 160 (22\%) & 127 (21\%) & 222 (33\%) \\
    $\Sigma$ True + Weak{-}ATD & 470 (70\%) & 155 (25\%) & 295 (44\%) \\
  \bottomrule
\end{tabular}
\end{table}

These results show that keyword-based filtering is useful, but not sufficiently precise to function as a standalone ATD detection method. A substantial proportion of issues flagged as ATD candidates were confirmed to be ATD-related, indicating that the extracted keywords capture relevant architectural signals. However, the relatively high false-positive rate indicates that lexical and semantic similarity alone are not sufficient to distinguish ATD from Non{-}ATD issues with similar wording.

To further examine false negatives, we randomly selected 384 presumed Non{-}ATD issues from each keyword-based method, resulting in a total of 1,152 Non{-}ATD samples. These were independently reviewed by two authors. Each Non{-}ATD instance required approximately 45 seconds to label, resulting in about 14 hours and 24 minutes of work per annotator. The issues were classified as ATD or Non{-}ATD, enabling computation of Cohen's kappa, which produced an agreement score of +0.81, indicating almost perfect agreement \cite{landis1977measurement}.

The Non{-}ATD validation results show that keyword-based methods were relatively effective at excluding clearly irrelevant issues, with approximately 83\%--85\% of sampled Non{-}ATD issues being correctly classified. Specifically, CS KeyBERT correctly identified 319 (83\%) of Non{-}ATD samples, while KeyBERT and TF-IDF correctly classified 328 (85\%) and 326 (85\%) items, respectively. This indicates that keyword-based filtering is cost-effective for reducing the search space, especially when ATD indicators are explicit and align well with the extracted keyword patterns.

At the same time, the remaining 15\%--17\% false negatives indicate that some ATD instances are expressed too implicitly or contextually to be captured through keyword matching alone. Taken together, the ATD-side and Non{-}ATD{-}side validation results indicate that keyword-based filtering is most effective as a triage mechanism for large-scale dataset construction rather than as a final classifier.

As a result of this combined evaluation process, including the manual validation of both ATD and Non{-}ATD samples, we expanded the initial ATD seed dataset of 57 expert-validated items to 1,100 ATD instances, including both True{-}ATD and Weak{-}ATD cases.

Keyword extraction provides a practical first-step mechanism for identifying candidate ATD issues at scale, but it does not achieve sufficient accuracy to serve as a standalone detection approach. Its main value lies in reducing the search space and enabling the construction of a large labeled dataset for downstream learning. These findings justify the next stage of the pipeline, in which active learning is used to improve detection performance while controlling annotation effort.

\begin{tcolorbox}[colback=gray!5!white, colframe=black, title=Summary (RQ1), boxrule=0.01pt, fonttitle=\small, fontupper=\small]
Keyword-based filtering is useful for expanding the dataset at scale, but not accurate enough for standalone ATD detection. Although it filters Non{-}ATD issues relatively well (83--85\%), its performance for identifying True{-}ATD cases remains limited (21--33\%). Thus, its main value lies in serving as an initial triage mechanism prior to supervised learning.
\end{tcolorbox}

The results of RQ1 indicate that keyword-based filtering is effective as a first-step triage mechanism for identifying candidate ATD issues, but it is insufficient as a standalone detection approach. This motivates the next stage of the pipeline, where we investigate whether active learning can improve ATD detection performance while reducing annotation effort.


\subsection{RQ2. Active Learning for ATD Detection}
\label{subsectionRQ2}

To answer RQ2, we evaluated whether active learning can improve ATD detection performance while reducing manual labeling effort. We compared six query strategies in a BERT-based active learning setting and then benchmarked the best-performing strategy against a range of traditional machine learning, deep learning, and transformer-based baselines. Results are reported for two labeling configurations: (i) using only True{-}ATD as the positive class and (ii) merging True{-}ATD and Weak{-}ATD into a single ATD class.

\subsubsection{Performance of Query Strategies}
Table~\ref{tab:query-strategy} reports the performance of the six query strategies under the two labeling configurations. Overall, the differences between strategies are modest in the True{-}ATD{-}only setting, but clearer gains emerge when Weak{-}ATD is included in the positive class.

\begin{table}[htpb!]
  \caption{Performance of AL query strategies under two ATD labeling configurations.} 
  \label{tab:query-strategy}
  \begin{tabular}{m{3.63cm}>{}m{0.9cm}>{}m{0.6cm}>{}m{1.1cm}>{}m{0.9cm}>{}m{0.6cm}>{}m{1.1cm}}
    \toprule
    \multirow{2}{9em}{Query Strategy} & \multicolumn{3}{c}{True{-}ATD Only} & \multicolumn{3}{c}{True- and Weak{-}ATD}\\
    \cline{2-7}
     & Precision & Recall & F1 score & Precision & Recall & F1 score \\
    \midrule
    Random & 0.64 & 0.65 & 0.64 & 0.68 &  0.68 & 0.68\\
    Least Confidence & 0.69 &  0.63 & 0.63 & 0.68 & 0.69 & 0.68\\
    Prediction Entropy & 0.65 & 0.65 & 0.65 & 0.71 & 0.70 & 0.70\\
    Embeddings K-Means & 0.64 & 0.65 & 0.63 & 0.70 & 0.70 & 0.70\\
    \rowcolor{lightgray}Breaking Ties & \textbf{0.68} & \textbf{0.68} & \textbf{0.68} & \textbf{0.72} & \textbf{0.71} & \textbf{0.72}\\
    Contrastive Active Learning & 0.66 & 0.64 & 0.65 & 0.70 & 0.72 & 0.70\\
  \bottomrule
\end{tabular}
\end{table}

\paragraph{\textbf{Performance on True{-}ATD Only.}}
When using only True{-}ATD labels, the active learning query strategies produced F1 scores within a relatively narrow range, from 0.63 to 0.68. This more modest performance is expected, as True{-}ATD cases are fewer and more difficult to learn from in isolation, limiting the diversity of positive training signals available to the model. In this setting, strategies that focus on model uncertainty appear more effective, as they help prioritize informative samples given limited labeling resources. As shown in Table~\ref{tab:query-strategy}, Breaking Ties achieved the highest F1 score (0.68), followed by Prediction Entropy and Contrastive Active Learning (both at 0.65), while Least Confidence and Embeddings K-Means recorded the lowest F1 scores (0.63). These results suggest that, although the performance differences are modest, uncertainty-based active learning provides a measurable advantage over random sampling for True{-}ATD detection in low-resource settings.

\paragraph{\textbf{Performance on True- and Weak{-}ATD Combined.}} In the more inclusive setting where Weak{-}ATD items are merged with True{-}ATD, performance improves across all strategies. This improvement is expected, as Weak{-}ATD items share semantic and structural similarities with True{-}ATD, thereby providing the model with more varied training signals. Such findings are also consistent with the established text classification literature, in which increasing both the diversity and the quantity of positive examples helps address class imbalance and enhances the model's ability to generalize \cite{diez2015diversity}. By incorporating Weak{-}ATD cases, even if they are less explicit, the model is able to capture a broader spectrum of ATD manifestations, thereby resulting in more robust and reliable detection. As shown in Table~\ref{tab:query-strategy}, the Breaking Ties strategy outperforms others with an F1 score of 0.72, followed closely by Contrastive Active Learning and Prediction Entropy (both at 0.70).

It is worth noting that the two labeling configurations already provide useful insight into the role of Weak{-}ATD in the detection task. As shown in Table~\ref{tab:query-strategy}, the best F1 score in the True{-}ATD-only setting is 0.68, whereas the configuration that merges True{-}ATD and Weak{-}ATD achieves a higher F1 score of 0.72. While indirect, this comparison suggests that including Weak{-}ATD in the positive class provides additional useful information for model learning and supports improved detection performance.

\subsubsection{Learning Behavior Across Annotation Iterations}

\begin{figure}[htp] 
    \centerline{\includegraphics[trim=1.0cm 1.0cm 0.1cm 0.5cm, clip, width=0.85\textwidth]{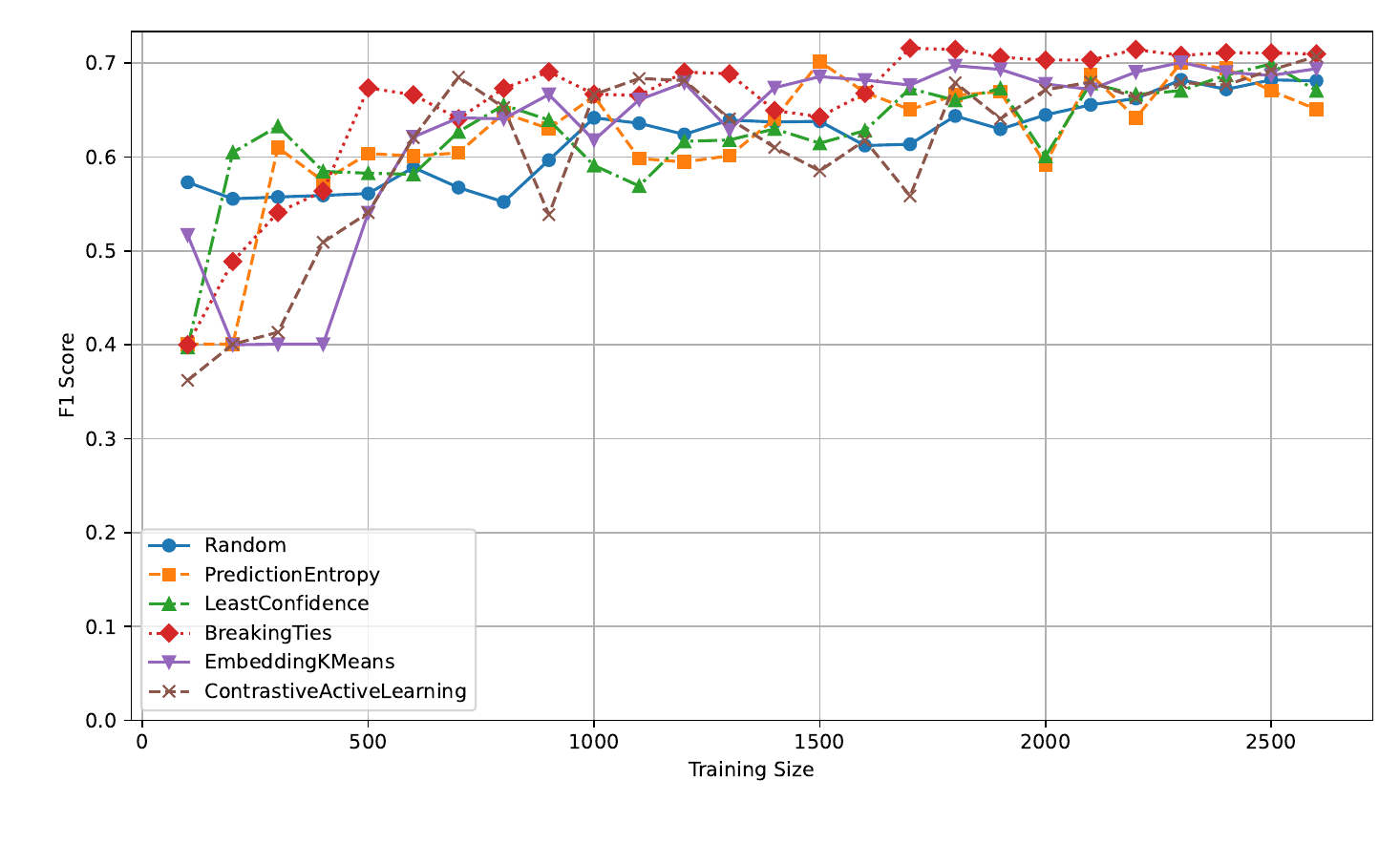}}
    \caption{F1 score trends of active learning query strategies across annotation iterations.}
    \label{fig_strategies}
 \end{figure}

Figure~\ref{fig_strategies} presents the performance comparison of the six query strategies used to detect ATD. The x-axis denotes the training size, while the y-axis shows the corresponding F1 score. For clarity and conciseness, we focus on reporting F1 scores in the main text, while the complete results, including precision and recall, are available in the replication package.

Prediction Entropy achieved its peak performance in iteration 14, with 1,500 samples (45\% of the dataset), achieving an F1 score of 0.70. In contrast, the Breaking Ties strategy exhibits a more consistent upward trend, achieving the highest overall F1 score of 0.72 in iteration 16 with 1,700 training instances (51\% of the dataset). Meanwhile, Contrastive Active Learning continued to improve steadily and achieved its best performance in iteration 25, with 2,600 labeled instances (78\%), resulting in an F1 score of 0.71.
Finally, Embedding K-Means achieved its highest F1 score of 0.70 at iteration 22, requiring 2,300 labeled samples (69\%). This shows that it performs competitively but requires a higher annotation budget than Prediction Entropy and Breaking Ties to achieve similar results.

The Random strategy showed moderate improvement across iterations, peaking at an F1 score of 0.68 in iteration 24 with 2,500 labeled instances (75\%), but never surpassing the best-performing uncertainty-based or diversity-based methods. Least Confidence, another uncertainty-based strategy, achieved a peak F1 score of 0.69 in iteration 15, using 1,600 training samples (48\%), outperforming Random but not as effectively as Breaking Ties or Prediction Entropy.

These results show that Breaking Ties is the most effective strategy for early performance, particularly with a relatively small labeled dataset, whereas Contrastive Active Learning demonstrates better scalability as training data increases. Overall, active learning strategies significantly improved the model's performance in identifying ATD by selecting informative samples during the labeling process.

Table~\ref{tab:results} presents the performance comparison of different models in detecting ATD, evaluated using precision, recall, and F1 score. Among all models, BERT with the Breaking Ties query strategy achieved the best overall performance, with an F1 score of 0.72, precision of 0.72, and recall of 0.71. This result indicates that integrating uncertainty-based active learning strategies, such as Breaking Ties, into a BERT-based classifier can significantly enhance ATD detection performance while reducing labeling effort.

\begin{table}[htp]
  \caption{Performance of baseline models and the best active learning for ATD detection.}
  \label{tab:results}
  \scriptsize
  \begin{tabular}{m{3.0cm}>{}m{2.2cm}>{}m{2.2cm}>{}m{2.2cm}}
    \toprule
    Model & Precision & Recall & F1 score \\
    \midrule
    \rowcolor{lightgray} BERT-Breaking Ties & \textbf{0.72} & \textbf{0.71}  & \textbf{0.72} \\
    BERT & 0.68 & 0.70 & 0.66 \\
    DeBERTa{-}v3 \cite{skryseth2023technical} & 0.68 & 0.69 & 0.68 \\
    RoBERTa \cite{sharma2022self} & 0.68 & 0.69 & 0.67\\
    ALBERT \cite{sharma2022self} & 0.67 & 0.68 & 0.67\\
    TextCNN \cite{li2020identification} & 0.67 & 0.70 & 0.67\\
    Text GCN & 0.61 & 0.60 & 0.61\\
    SVM & 0.64 & 0.61 & 0.62 \\
    LR & 0.67 & 0.70 & 0.65 \\
    NB & 0.67 & 0.50 & 0.41\\
    RF & 0.68 & 0.60 & 0.59\\
    kNN & 0.62 & 0.60 & 0.61 \\
    ME & 0.67 & 0.66 & 0.65 \\

  \bottomrule
\end{tabular}
\end{table}

In comparison, the baseline BERT model, trained on the entire labeled dataset without active learning, achieved slightly lower results: precision = 0.68, recall = 0.70, and F1 score = 0.66. This demonstrates that BERT-Breaking Ties, trained on only 51\% of the dataset, outperformed the baseline BERT (precision = 0.72, recall = 0.71, and F1 score = 0.72), highlighting the effectiveness of active learning in both reducing annotation costs and improving model performance.

Other transformer-based models also yielded competitive results. DeBERTa{-}v3 achieved a precision of 0.68, recall of 0.69, and F1 score of 0.68, slightly outperforming the baseline BERT in terms of F1 score. RoBERTa and ALBERT both obtained F1 scores of 0.67, demonstrating their capability but still falling short of BERT-Breaking Ties. The TextCNN model also performed well, achieving a precision of 0.67, a recall of 0.70, and an F1 score of 0.67.

Among other deep learning models, Text GCN lagged behind with an F1 score of 0.61, suggesting that graph-based document modeling may be less effective for this specific ATD detection task.

Traditional machine learning models consistently exhibited lower performance than transformer-based approaches. LR and ME achieved F1 scores of 0.67 and 0.65, respectively, with LR being the strongest among them. SVM, kNN, and RF yielded F1 scores of 0.62, 0.61, and 0.59, respectively. NB had the lowest recall (0.50) and F1 score (0.41), despite reasonable precision (0.67), indicating limited suitability for ATD classification.

In summary, the results reinforce that BERT-Breaking Ties not only achieves the highest accuracy but also reduces the annotation effort by nearly half. These findings highlight the value of active learning strategies in domains where labeled data is scarce or expensive to obtain, such as ATD detection.

\begin{tcolorbox}[colback=gray!5!white, colframe=black, title=Summary (RQ2), boxrule=0.01pt, fonttitle=\small, fontupper=\small]
The Breaking Ties active learning strategy consistently improves model performance, achieving the highest F1 score of 0.72 at iteration 16 using only 1,700 labeled instances (51\% of the dataset). This demonstrates that active learning can significantly reduce annotation effort while enabling the BERT-based model to outperform conventional approaches.
\end{tcolorbox}

After establishing the classification performance of the proposed active learning approach, we will examine whether the resulting predictions can be meaningfully and usefully explained to domain experts.

\subsection{RQ3. Explainability and Expert Evaluation}
\label{subsectionRQ3}
To answer RQ3, we examined whether LIME and SHAP can provide useful and interpretable explanations for the BERT-based ATD classifier. We first illustrate the explanations produced by both methods on representative Jira issues, and then report the results of an expert evaluation study involving ten researchers with experience in technical debt, SATD, or software architecture.

\subsubsection{Illustrative Explanations from LIME and SHAP}

In our study, LIME was first used to generate local explanations for individual predictions made by the BERT-based ATD classifier. For each Jira issue, LIME outputs a ranked list of tokens (words or phrases) along with their contribution weights, indicating the degree to which each token influenced the model's classification as ATD or Non{-}ATD.

\begin{figure*}[htpb] 
    \centerline{\includegraphics[width=0.85\textwidth]{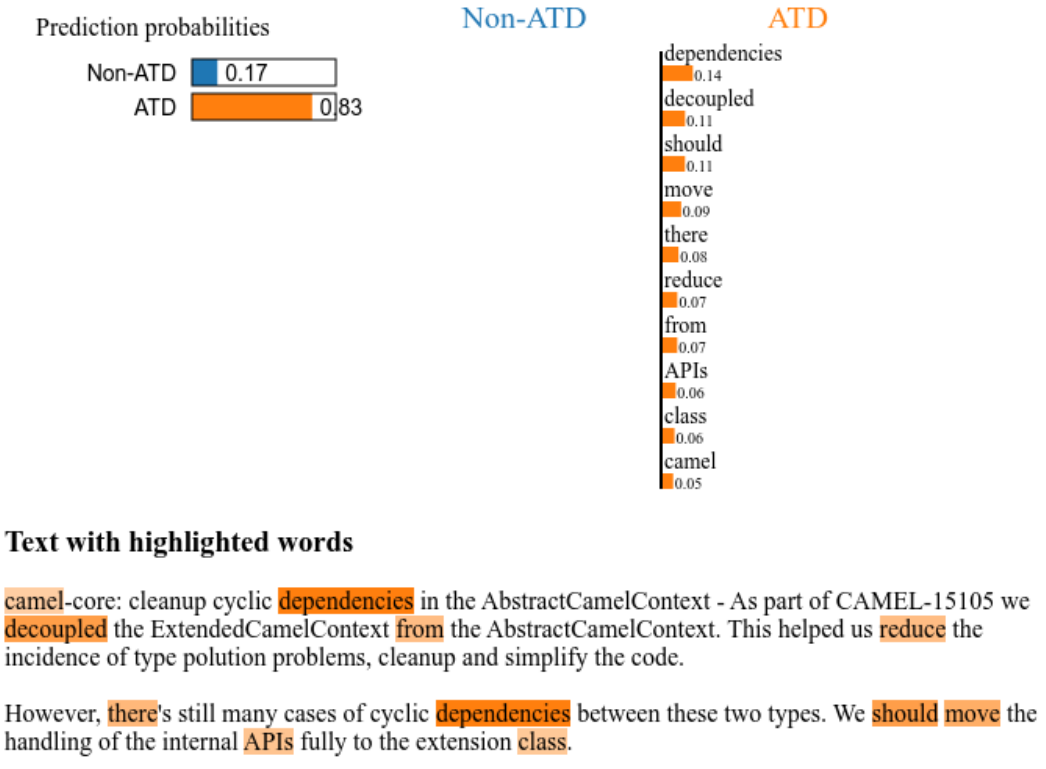}}
    \caption{LIME plot explanation for a Jira issue (CAMEL-19998) classified as ATD.}
    \label{fig_explain_lime-camel}
 \end{figure*}

Figure~\ref{fig_explain_lime-camel} illustrates a LIME explanation for one issue classified as ATD. The horizontal bar chart highlights the most influential tokens from the issue's summary and description. Tokens such as \enquote{dependency,} \enquote{decoupled,} \enquote{should move,} \enquote{refactor,} and \enquote{reduce} appear with positive weights, meaning they strongly support the model's decision to classify the issue as ATD. These terms reflect common architectural concerns, including modularity, abstraction, and dependency restructuring.

\begin{figure*}[htpb] 
    \centerline{\includegraphics[width=0.85\textwidth]{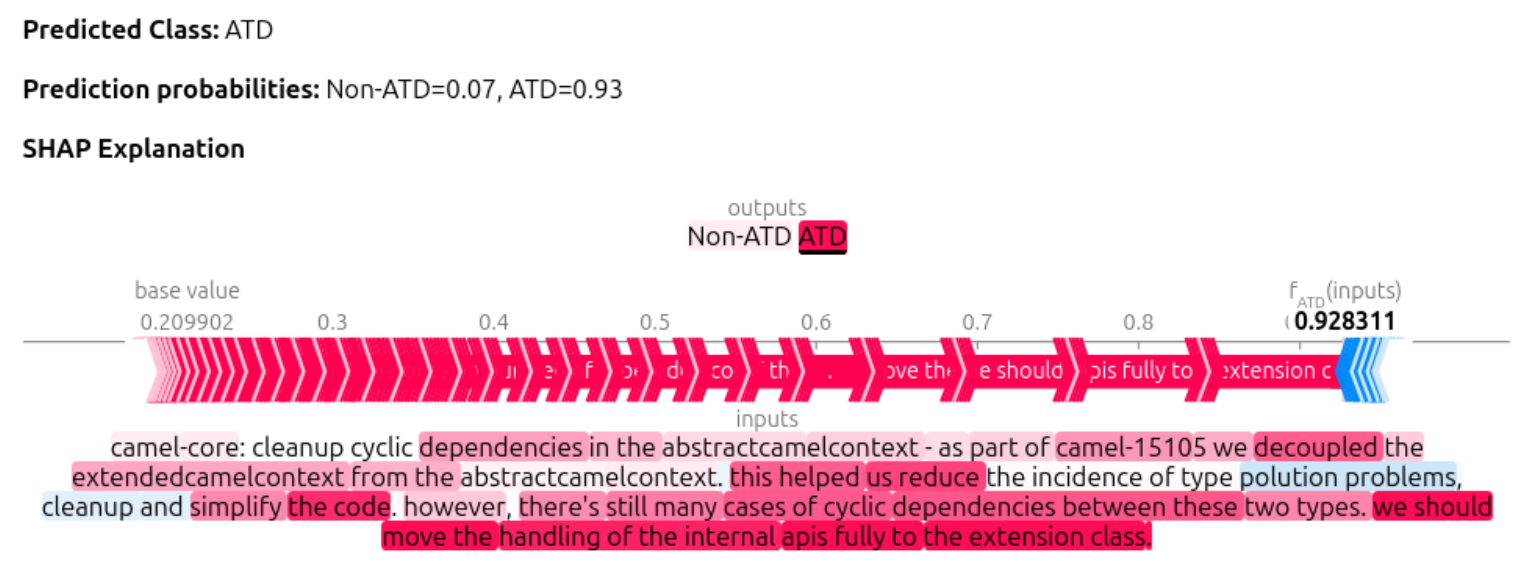}}
    \caption{SHAP plot explanation for a Jira issue (CAMEL-19998) classified as ATD.}
    \label{fig_explain_shap-camel}
 \end{figure*}

To further interpret the predictions made by our BERT-based classifier for ATD, we applied SHAP to generate local explanations for individual instances. Figure~\ref{fig_explain_shap-camel} displays a SHAP text explanation for a Jira issue classified as ATD. At the top of the figure, the predicted class (ATD) and the associated prediction probabilities (Non{-}ATD = 0.07, ATD = 0.93) are reported, providing transparency into the model's confidence.

The SHAP visualization highlights the specific tokens in the Jira issue text that most influenced the classification. Tokens highlighted in red contributed positively toward the ATD class, while tokens in blue (if present) would indicate contributions toward the Non{-}ATD class. The horizontal axis represents the model's output probability, starting from the base value (0.21) and moving toward the final predicted probability for the ATD class (0.93). The magnitude of each token's contribution is visually represented by the length and intensity of its highlight.

In this example, tokens such as \enquote{decoupled}, \enquote{simplify the code}, and \enquote{should move} are prominently highlighted in red, signaling strong support for the ATD prediction. This local explanation helps users and experts see which parts of the text the model considered most indicative of architectural technical debt, supporting transparency and interpretability in automated ATD detection.

\subsubsection{Qualitative Feedback from Experts}
While these explainability tools provide a technical breakdown of model behavior, the ultimate usefulness of such explanations depends on whether they are meaningful to human stakeholders, particularly technical debt experts. To assess the practical usefulness and interpretability of these explanations, we conducted an evaluation study with SE researchers experienced in technical debt and software architecture. Each expert was asked to review model predictions along with their corresponding explanations generated by LIME and SHAP and to rate them across eight attributes adapted from Al{-}Ansari et al. \cite{al2024user}. The primary objective of this evaluation was to determine whether the highlighted tokens and the underlying reasoning provided by these XAI methods aligned with experts' understanding of architectural technical debt. By collecting structured feedback on these aspects, we aimed to identify the key characteristics that constitute effective explanations for ATD classifications, as determined by expert judgment.

We invited 17 expert researchers, each with at least five years of experience in TD, SATD, or software architecture, to participate in our evaluation study. Ten of these researchers responded and participated in the study. These ten experts reviewed representative ATD and Non{-}ATD predictions, along with the corresponding explanations generated by LIME and SHAP. Their feedback was collected through a structured questionnaire that assessed eight key attributes.


The collected responses provide valuable insights into how experts perceive the quality and usefulness of XAI explanations in the context of ATD detection. Table~\ref{tab:survey-result} summarizes the mean scores assigned to each XAI goal for both LIME and SHAP, revealing important differences in how these methods are evaluated across the eight attributes. Overall, both explanation methods received moderate to high scores, suggesting that they are considered valuable tools for enhancing the interpretability of ATD classification. However, their mean ratings highlight distinct strengths and limitations for each technique.

\begin{table}[htp]
  \caption{Summarizes the mean scores for each XAI goal across LIME and SHAP.} 
  \label{tab:survey-result}
  \scriptsize
  \begin{tabular}{m{3.0cm}>{}m{2.2cm}>{}m{2.2cm}}
    \toprule
    XAI Goal & LIME & SHAP \\
    \midrule
    Trustworthiness & \textbf{3.40} & 3.10\\
    Informativeness & 3.30 & \textbf{3.50}\\
    Interactivity & \textbf{3.50} & 3.30\\
    Fairness & \textbf{3.80} & 3.20\\
    Confidence & \textbf{3.70} & 3.50\\
    Accessibility & \textbf{4.00} & 2.40\\
    Causality & \textbf{4.10} & 3.10\\
    Transferability & \textbf{3.50} & 3.10\\ 
  \bottomrule
\end{tabular}
\end{table}

The mean scores for each XAI goal were rated on a five-point scale. In terms of \textbf{trustworthiness}, LIME received a mean score of 3.40, while SHAP received 3.10. For \textbf{informativeness}, SHAP was rated slightly higher (3.50) than LIME (3.30). LIME also obtained higher scores than SHAP for \textbf{interactivity} (3.50 vs. 3.30) and \textbf{accessibility} (4.00 vs. 2.40). 

With respect to \textbf{fairness}, LIME scored 3.80 compared to SHAP's 3.20. For \textbf{confidence}, LIME received 3.70 and SHAP 3.50. In terms of \textbf{causality}, LIME scored 4.10, while SHAP scored 3.10. For \textbf{transferability}, LIME received a mean score of 3.50 and SHAP 3.10.

In addition to the quantitative results, qualitative feedback from participating experts provides valuable insights into their experiences with LIME and SHAP explanations. The qualitative data generally corroborate the trends observed in Table~\ref{tab:survey-result}, revealing important nuances in user perception. 

For both LIME and SHAP, experts emphasized that the usefulness of the explanations often depended on the relevance of the highlighted features. Several experts appreciated the straightforward visualizations provided by LIME. As one expert noted, \enquote{\textit{The bar charts and highlighted text show why the model may have made the decision, which would help someone without the expertise.}} These visual aids facilitated the identification of influential features and supported a clearer understanding of causal relationships. This qualitative feedback aligns with the higher scores achieved by LIME across almost all XAI goals.

\begin{figure*}[htpb] 
    \centerline{\includegraphics[trim=0.3cm 2.1cm 0.1cm 1.75cm, clip, width=0.35\textwidth]{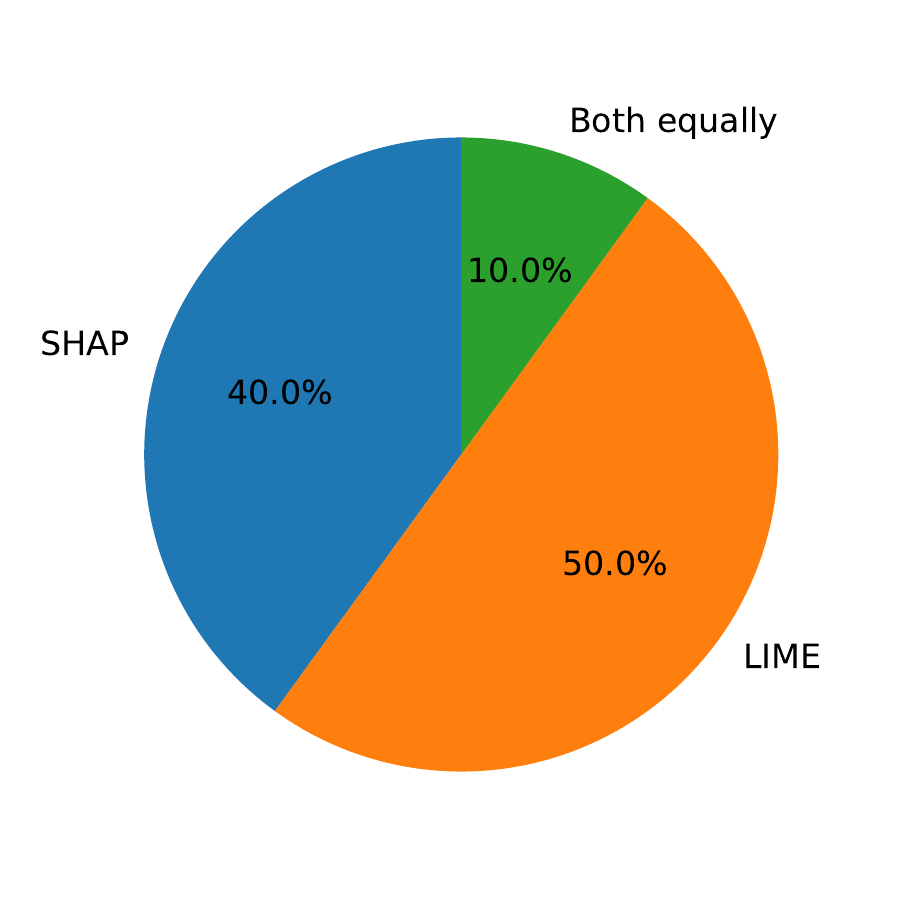}}
    \caption{XAI Method Preference.}
    \label{fig:xai-preference}
 \end{figure*}

SHAP was recognized for producing more detailed and theoretically grounded outputs, although expert opinions varied regarding accessibility. Several experts appreciated the comprehensive breakdowns offered by SHAP and the clear highlighting of influential phrases, noting that these features were helpful for interpreting model predictions and assessing confidence. This perspective helps explain the higher score achieved by SHAP for informativeness, as experts considered its explanations to provide richer details about model reasoning. For example, one expert commented, \enquote{\textit{I think the SHAP plot is much more confusing to someone who may not have expertise in machine learning compared to the LIME plot, but the highlighted text using SHAP gives better insight into the model's decisions compared to LIME}}. At the same time, many found SHAP less accessible or more challenging to interpret, which is consistent with the consistently higher mean scores obtained by LIME. 

When asked about their overall preference for XAI methods (see Figure~\ref{fig:xai-preference}), 50\% of the experts favored LIME, citing its clarity and ease of use. Another 40\% preferred SHAP, highlighting its depth and richness of information. The remaining 10\% found both methods equally effective. Several experts also suggested improvements to both techniques, including filtering out irrelevant features and adding interactive elements to enhance comprehension and trust in the explanations.

In summary, the qualitative findings reinforce and extend the quantitative results. While both LIME and SHAP are valuable for supporting expert interpretation of ATD classification, each method presents distinct strengths and trade-offs. These insights suggest that the choice between LIME and SHAP may ultimately depend on the specific needs and expertise of the intended audience and the complexity of the explanation task.


\begin{tcolorbox}[colback=gray!5!white, colframe=black, title=Summary (RQ3), boxrule=0.01pt, fonttitle=\small, fontupper=\small]
LIME and SHAP both improved the interpretability of ATD classification results for experts. LIME was rated higher across almost all XAI goals, while SHAP performed slightly better in terms of informativeness. Both methods provided valuable support for expert validation, highlighting the usefulness of XAI techniques in ATD detection.

\end{tcolorbox}

\section{Discussion}
\label{sectionDiscussion}
This study explored a hybrid approach to reducing annotation effort for detecting ATD in Jira issue tracking systems by combining keyword-based filtering, active learning, and XAI. Below, we discuss the implications of our findings with respect to the research questions and the broader context of ATD management and AI adoption in SE.

\subsection{Effectiveness of Keyword-based Filtering}
The results of RQ1 in Section~\ref{subsectionRQ1} indicate that keyword-based filtering is highly effective for removing Non{-}ATD issues from large datasets, as shown by accuracy rates between 83\% and 85\%. This outcome demonstrates that targeted keyword selection, when grounded in expert-validated ATD examples, can serve as an efficient and scalable approach for initial data reduction. By filtering out most irrelevant issues, keyword-based methods significantly decrease the manual effort required in subsequent annotation phases.

However, the recall for identifying True{-}ATD issues remains low, with only 21\% to 33\% of such cases successfully detected. This finding highlights a critical limitation: keyword-based filtering alone is not sufficient to capture the full range of ATD expressions in issue trackers. This result is consistent with findings from Maldonado et al.~\cite{da2017using}, who reported that keyword patterns also achieved low recall when identifying SATD in code comments. 

Many ATD instances are articulated through nuanced or project-specific language, which frequently falls outside the boundaries of even the most carefully constructed keyword lists. As a result, a substantial proportion of actual ATD cases remain undetected when relying solely on keyword-based techniques. Additionally, the observed high false positive rates suggest that there is considerable lexical overlap between ATD and Non{-}ATD issues, complicating the task of accurate identification using simple lexical cues alone.

These findings indicate that while keyword-based filtering is valuable for narrowing down the annotation workload, it cannot be considered a standalone solution for robust ATD detection. Its primary strength lies in serving as an initial triage mechanism, making large-scale annotation projects more tractable and cost-efficient. To achieve comprehensive and precise identification of architectural technical debt, it is essential to integrate keyword-based filtering with more advanced methods, such as supervised learning models or semantic analysis techniques. These approaches can leverage context and capture implicit indicators of ATD that are not accessible through keywords alone.

\paragraph{\textbf{Implication for Researchers.}} Our findings suggest that keyword-based filtering alone is insufficient for robust ATD identification, although it remains valuable as an initial triaging mechanism that reduces the manual inspection space before applying supervised models. More broadly, the proposed pipeline may be applicable beyond ATD and object-oriented software systems. It may also support the identification of technical debt in other underexplored and domain-specific software contexts, such as quantum software, machine learning-based software, deep learning frameworks, scientific software, and blockchain projects, where labeled data are often scarce and textual signals are highly context-dependent. In such settings, combining keyword-based candidate generation, active learning, and XAI may offer a practical strategy for making annotation and model development more feasible.


\paragraph{\textbf{Implication for Practitioners.}} Keyword-based filtering can serve as a first-line tool to significantly reduce manual annotation workloads, helping practitioners quickly focus their attention on potential ATD issues. However, practitioners should combine this initial filter with expert reviews or machine-assisted validation for the best results. Given the risk of false positives, practitioners should be cautious about relying solely on keyword-based approaches. It is recommended to incorporate this approach into broader ATD detection strategies rather than using it as a standalone solution.

\subsection{Active Learning for Efficient ATD Detection}
The results of RQ2 (as shown in~\ref{subsectionRQ2}) provide clear evidence that active learning is a highly effective strategy for reducing annotation effort while maintaining strong ATD detection performance. Among the evaluated query strategies, Breaking Ties consistently delivered the best results, achieving an F1 score of 0.72 while requiring labels for only 51\% of the dataset. Although the improvement over the baseline BERT model is moderate in absolute terms, the result remains practically meaningful, as it demonstrates that comparable or better performance can be achieved with substantially fewer labeled instances. In the context of ATD detection, where annotation requires architectural expertise and is costly to obtain, this reduction in labeling effort is an important benefit.

The comparison across query strategies further suggests that active learning can make better use of limited annotation budgets than random sampling. In particular, the strong performance of Breaking Ties indicates that prioritizing instances with small margins between the two most probable classes can be effective for ATD detection, where issue reports often contain implicit or borderline architectural signals. Rather than dramatically increasing predictive performance, active learning appears to improve the efficiency with which informative training examples are selected. This is consistent with prior work showing that active learning can be valuable in software engineering tasks where labeled data is scarce or expensive to obtain~\cite{dor2020active,miller2020active,jacobs2021active}. 

At the same time, these gains should be interpreted with appropriate caution from a deployment perspective. Pool-based active learning requires repeated retraining of the classifier across annotation rounds, and using BERT as the base learner introduces non-trivial computational overhead. As a result, the benefit of reduced human labeling effort must be balanced against the cost of iterative model retraining. In research settings or low-resource ATD scenarios, where expert annotation is the primary bottleneck, this trade-off may still be favorable. However, in practical deployment settings with limited computational resources or strict time constraints, repeated BERT retraining may reduce the operational advantage of the approach. Future work should therefore investigate more deployment-oriented alternatives, such as warm-start fine-tuning, less frequent model updates, or lighter base learners.

These findings reinforce the importance of efficient annotation strategies for ATD, which has long been recognized as one of the most costly and consequential forms of technical debt in software systems~\cite{martini2015danger,lenarduzzi2019technical}. While prior studies have investigated technical debt detection more broadly, fewer have focused specifically on reducing labeling effort for ATD identification. Our results suggest that active learning is a promising direction for this setting, not because it yields a dramatic performance increase, but because it helps make ATD dataset construction more feasible and scalable.

\paragraph{\textbf{Implication for Researchers.}} These results suggest that active learning is worth further investigation for software engineering tasks in which labels are scarce, expensive, or require domain expertise. In particular, the effectiveness of Breaking Ties indicates that uncertainty-based sampling is a strong candidate for ATD-related classification problems. At the same time, the moderate absolute performance gains and the computational cost of iterative retraining indicate that future work should not focus only on predictive performance, but also on the broader cost-benefit trade-off between labeling effort and computational effort. This opens opportunities to study more efficient active learning workflows, alternative base learners, and to treat True{-}ATD and Weak{-}ATD as distinct learning targets.

\paragraph{\textbf{Implication for Practitioners.}} For practitioners, the results suggest that active learning can be useful when expert annotation is expensive and only a limited labeling budget is available. In such cases, query strategies such as Breaking Ties may help prioritize the most informative issues for review and support more efficient ATD dataset construction. However, the practical adoption of this approach should also consider computational constraints, since repeated retraining of transformer-based models may not always be cost-effective in operational environments. The method is therefore most attractive in settings where reducing expert labeling effort is more critical than minimizing model training cost.

\subsection{Explainability in ATD Classification}
The findings from RQ3 underscore the critical role of XAI techniques in supporting the adoption and practical impact of automated ATD detection. By applying LIME and SHAP to the predictions of the BERT-based classifier, the study demonstrates that model transparency can be substantially enhanced, enabling better understanding of the factors driving ATD classification outcomes. These explanation tools enable users to identify which words, phrases, or linguistic features most strongly influenced the model's decisions, addressing a key challenge in deploying transformer-based models for technical debt management.

Feedback from the expert evaluation reveals that explanations generated by LIME and SHAP are perceived as valuable for supporting interpretability, trust, and actionable insight. Experts expressed a clear preference for LIME over SHAP, particularly regarding accessibility and interpretability. While SHAP was recognized for producing more detailed and theoretically grounded outputs, many experts found LIME's explanations more intuitive and straightforward. Several noted that LIME's visualizations and clear identification of influential features facilitated rapid inspection and understanding of model predictions. This preference is reflected in LIME's consistently higher mean scores across most evaluation criteria. The result is also consistent with the findings of Jiarpakdee et al. \cite{jiarpakdee2021practitioners}, who identified LIME as the most preferred technique for understanding the key characteristics contributing to a prediction.

Although SHAP received a higher informativeness score due to its comprehensive breakdowns and richer details about model reasoning, its explanations were often considered less accessible and more challenging to interpret than those of LIME. This finding aligns with concerns raised by Kumar et al. \cite{kumar2020problems}, who argued that explanations generated by techniques such as SHAP are not always easy to understand and often do not align with human expectations. In practice, users tend to prefer technically accurate, intuitive, and meaningful explanations from a human perspective.

It is worth noting, however, that some highlighted tokens, such as \enquote{from,} \enquote{class,} and \enquote{camel,} may be less informative or even irrelevant for ATD detection. This limitation is inherent to current explanation methods, which sometimes assign importance to common or project-specific terms. Such occurrences indicate the need for further refinement of feature selection or filtering strategies in future work to improve the practical value and interpretability of model explanations.

\paragraph{\textbf{Implication for Researchers.}} The variance in expert preferences and the differing evaluations of SHAP and LIME regarding accessibility, informativeness, and interpretability highlight the need for further research into domain{-}specific customization of explainability methods. Researchers should explore how tailored explainability tools specifically designed for ATD contexts might better align with expert expectations and provide deeper insights into model reasoning. To support this, future research should also aim to establish standardized evaluation frameworks or benchmarks, including clear criteria and standards, to enable objective comparisons and help practitioners select the most appropriate explainability methods for their contexts.


\paragraph{\textbf{Implication for Practitioners.}} Given the clear expert preference for LIME, practitioners implementing automated ATD detection should prioritize explainability tools that provide accessible and intuitive visual explanations. LIME's visualizations can effectively bridge communication gaps among stakeholders with varying levels of technical expertise, fostering trust and enabling more informed decisions. However, practitioners should also consider their audience's familiarity with technical details when selecting explainability tools. While SHAP was perceived as more challenging to interpret, it offers deeper, more comprehensive insights that can be particularly valuable for technical teams or in scenarios requiring detailed justification of model reasoning. Thus, practitioners might adopt a strategic approach, employing LIME for broad stakeholder communication and SHAP for cases that demand in-depth technical analysis.

\subsection{Threats to Validity}
\paragraph{\textbf{Construct Validity: }} A potential threat to construct validity arises from the way architectural technical debt is operationalized in this study. First, the proposed approach identifies self{-}admitted ATD from Jira issue text and therefore detects textual evidence of architectural debt discussions rather than all possible manifestations of ATD in the system. In this sense, detecting technical debt through textual artifacts can serve as a complementary approach to existing code smell detectors based on source code analysis~\cite{da2017using}. This limitation is important because architectural debt is inherently structural, and some architecturally relevant problems may never be explicitly documented in issue reports. Consequently, such instances would remain invisible to our approach.

A second threat arises from merging True{-}ATD and Weak{-}ATD into a single \enquote{ATD} class during model training and evaluation. While this decision simplifies the classification task and helps address data sparsity, it introduces some ambiguity into the conceptual boundary of the ATD construct. This choice aligns with the main goal of the study, namely, reducing labeling effort for ATD detection, rather than performing a fine-grained classification of ATD subtypes. To mitigate these threats, we grounded the annotation process in established ATD indicators, used multiple annotators with majority voting to reduce subjectivity, and explicitly tracked Weak{-}ATD items to maintain transparency in class composition. Although we merged the classes for this initial classification task to improve training stability, we preserved the distinction between True{-}ATD and Weak{-}ATD in the replication package, enabling future studies to revisit the problem using multi-class or hierarchical modeling approaches. Future work should also complement textual issue analysis with additional architectural sources, such as dependency graphs, architecture decision records, and design documentation, to obtain a more complete view of ATD.

\paragraph{\textbf{Internal Validity:}} Several design choices in the pipeline may have influenced the reported results. First, the seed dataset used for keyword extraction was intentionally conservative, retaining only issues unanimously labeled as ATD by all three authors. Although this improves seed reliability, it may bias keyword extraction toward clearer and more explicit ATD expressions while underrepresenting subtler architectural cases. Second, the performance of keyword-based filtering may be affected by project-specific terminology and by the semantic behavior of the extraction methods. We mitigated this risk by combining statistical and semantic keyword extraction approaches, manually validating representative samples, and comparing multiple extraction methods. Third, in the active learning experiments, we evaluated several query strategies but used BERT as the base learner for the full iterative setup. This choice was made to balance effectiveness and computational feasibility, but it means that the interaction between query strategies and other transformer architectures remains underexplored. Finally, although active learning improved the trade-off between labeling effort and performance, the gains over the supervised baseline were moderate. Therefore, the practical value of the approach depends on settings where expert labeling cost is a more severe bottleneck than iterative model retraining. Future work should investigate warm-start training, other transformer architectures, and ablation studies on seed selection and project composition.

\paragraph{\textbf{External Validity: }} Since our dataset is derived solely from Apache open-source Java projects using Jira, the results may not fully generalize to industrial software, other programming languages, alternative issue trackers such as GitHub, or non-Apache ecosystems. In particular, Apache projects may share similar process conventions, reporting culture, and issue-template characteristics, which could make the linguistic structure of Jira issues in our dataset more homogeneous than in other contexts. This homogeneity may limit the generalizability of both the learned keyword patterns and the classification models. To reduce this threat, we selected 10 large, diverse, and actively maintained projects spanning multiple domains. Future work should extend this approach to other issue tracking systems, programming ecosystems such as Python and C++, industrial datasets, and additional architectural sources, including design documents, architecture decision records, and dependency graphs, to improve cross-context generalizability.

In addition, the evaluation study on XAI in ATD detection employs a systematic sampling approach, as suggested by Stol and Fitzgerald \cite{stol2018abc}. Participants are selected based on their expertise in relevant research domains rather than using a representative sample of the broader practitioner or developer population. As a result, while the findings provide in-depth, informed judgments from domain experts, they may not be directly generalizable to all practitioners or to the broader SE community. This is appropriate given that the primary objective is to obtain informed judgment rather than to generalize the results to the entire population. Nevertheless, examining practitioners' perspectives using a representative sampling approach would be a valuable direction for future work.

\paragraph{\textbf{Reliability: }} 
In this study, the selection of seed keywords for CS KeyBERT was based on prior literature and researcher judgment, which may have shaped the direction of keyword extraction toward expected patterns, potentially excluding alternative formulations of architectural technical debt. To improve reliability, the selection of seed keywords were extracted from their high frequency and strong association with ATD from the previous dataset. However, we acknowledge that using an expert panel or practitioner input in future work would further enhance objectivity.

\section{Conclusion and Future Work}
\label{sectionConclusion}
In this study, we introduced a new approach to annotating and detecting ATD in Jira issue tracking systems by combining keyword-based methods with active learning. This work was motivated by the scarcity of labeled ATD instances, which makes it difficult to train effective machine learning models and hinders progress in automated technical debt detection.

Our findings indicate that keyword-based filtering, although limited in recall, can identify only 21 to 33\% of True{-}ATD items but can significantly reduce the manual labeling workload, filtering out up to 85\% of Non{-}ATD issues. To address the limitations of keyword approaches, we applied active learning strategies that focus annotation efforts on the most informative samples. This enabled our BERT-based classifier to achieve an F1 score of 0.72 while only requiring half of the available labeled data.

Our comparative analysis revealed that the Breaking Ties query strategy delivers the best performance in the early annotation stages, while Contrastive Active Learning is more effective for larger annotation budgets. By combining lightweight keyword filtering with active learning, we propose a practical and cost-efficient method for building robust ATD datasets and detection models.

To improve the interpretability of model predictions, we integrated XAI techniques, such as LIME and SHAP, to enhance their transparency. To assess the effectiveness of these methods, we conducted an expert evaluation involving ten experts. The results demonstrated that both LIME and SHAP significantly improved practitioners' understanding and trust in automated ATD classification results. Notably, most experts preferred LIME over SHAP, citing its more intuitive and accessible explanations. However, the evaluation also indicated that further customization of these methods may be beneficial to better address the specific needs of architectural contexts.

To improve reproducibility and support future research, we have made our dataset publicly available, which includes a new set of 1,100 ATD items. The workflow we propose offers a scalable solution for identifying technical debt in large software projects, helping teams focus on issues with high architectural impact.

Looking ahead, future research should investigate ATD identification across multiple sources, expand to additional software domains and artifact types, and explore integrating ATD detection into real{-}time development workflows. Further studies on human{-}centered evaluation of XAI will also be important for advancing technical debt management in practice.

\section*{Acknowledgment}
This work was financially supported by the Indonesian Education Scholarship, Center for Higher Education Funding and Assessment, and Indonesia Endowment Fund for Education.

\section*{Author Contributions}
\textbf{Edi Sutoyo}: Conceptualization; Data Curation; Data Labeling; Methodology; Writing—Original Draft; Visualization. \textbf{Paris Avgeriou}: Conceptualization; Data Labelling; Methodology; Supervision; Writing—Review \& Editing. \textbf{Andrea Capiluppi}: Conceptualization; Data Labeling; Methodology; Supervision; Writing—Review \& Editing.

\section*{Data Availability}
The data associated with this study are publicly available online in the replication package.\footnote{\url{https://github.com/edisutoyo/atd-issues}}

\section*{Declarations}

\paragraph{\textbf{Conflicts of interests}\\}
The authors have no competing interests to declare that are relevant to the content of this article.

\paragraph{\textbf{Ethical approval}\\}
Not applicable.

\paragraph{\textbf{Informed Consent}\\}
Not applicable.

\paragraph{\textbf{Clinical Trial Number}\\}
Not applicable.

\paragraph{\textbf{Funding}\\}
This work was financially supported by the Indonesian Education Scholarship, Center for Higher Education Funding and Assessment, and Indonesia Endowment Fund for Education.


\tiny
\bibliographystyle{spmpsci}
\bibliography{Main}

\appendix









\end{document}